\documentclass{article}[12pt]

\usepackage{lmodern}
\usepackage[english]{babel}
\usepackage{amsmath}
\usepackage{graphicx}
\usepackage[colorlinks=true, allcolors=blue]{hyperref}
\usepackage{bm}
\usepackage{extarrows}
\usepackage{algorithm}
\usepackage{algpseudocode}
\usepackage{booktabs}
\usepackage{xcolor}
\usepackage{ragged2e}
\usepackage{makecell}
\usepackage{multirow}
\usepackage{multicol}
\usepackage{tabularx}
\usepackage{colortbl}
\usepackage{natbib}
\usepackage{diagbox}
\usepackage{subcaption}
\usepackage[title]{appendix}

\definecolor{brown}{rgb}{0.8, 0.33, 0.1}

\DeclareMathOperator*{\argmax}{argmax}
\DeclareMathOperator*{\argmin}{argmin}

\newcommand{\logit}{\text{logit}}
\newcommand{\emax}{E_{\text{max}}}
\newcommand{\edft}{ED_{50}} 
\newcommand{\thetaPK}{\bm{\theta}_{\text{PK}}}

{
	\title{\bf   Pharmacometrics-Enabled    DOse OPtimization (PEDOOP) for Seamless Phase I-II Trials in Oncology}

 	\author{
 		Shijie Yuan\thanks{Department of Statistics and Data Science, The University of Texas at Austin, Austin, USA} , Zhanbo Huang\thanks{School of Data Science, Fudan University, Shanghai, CHN} , Jiaxin Liu\thanks{Cytel Inc., Shanghai, CHN} , and Yuan Ji\thanks{Department of Public Health Sciences, The University of Chicago, Chicago, USA; Corresponding email: koaeraser@gmail.com}
 	}
	\date{\today}
}

\begin{document}
\maketitle

\begin{abstract}
We consider a dose-optimization design for   first-in-human oncology trial that   aims    to identify a suitable dose for late-phase drug development. The proposed approach, called the Pharmacometrics-Enabled DOse OPtimization (PEDOOP) design,  incorporates observed patient-level pharmacokinetics (PK) measurements and latent pharmacodynamics (PD) information for trial decision making and dose optimization.  PEDOOP consists of two   seamless    phases.    In phase I, patient-level time-course drug concentrations,   derived    PD effects, and the toxicity outcomes from patients are integrated into a statistical model to estimate the   dose-toxicity response.    A simple dose-finding design guides dose escalation in phase I.  At the end of the phase I dose finding,   a graduation rule is used to  assess the safety and efficacy of all the doses and select those with promising efficacy and acceptable safety  for a randomized comparison  against a control arm  in phase II. In phase II, patients are randomized to the selected doses   based on a  fixed or adaptive randomization ratio. At the end of phase II, an   optimal biological dose (OBD)    is selected for late-phase development. We conduct simulation studies to assess the PEDOOP design in comparison to an existing seamless design that also combines phases I and II in a single trial. 

\end{abstract}


\section{Introduction}
   Early-phase  oncology trials like phase I and phase  II  studies investigate the  toxicity and efficacy of a new drug.    In most phase I   trials,    the goal is to explore the toxicity profile of the tested drug and identify the maximum tolerated dose (MTD), the highest dose that does not cause unacceptable side effects. If a target toxicity rate $p_T$ is specified in the trial, the MTD is   defined    %represented 
as the highest dose with a toxicity probability less than $p_T$.   A large number of designs have been proposed for phase I trials, ranging from model-free designs like  the ``3+3" design \citep{storer1989design}  and the ``i3+3" design \citep{liu2020i3+}, model-based designs like the continual reassessment method (CRM) \citep{o1990continual}, the mTPI design \citep{ji2010modified}, the mTPI-2 designs \citep{guo2017bayesian}, and the BOIN design \citep{yuan2015boin}. Some model-based designs use independent likelihood and prior distributions, resulting  in   independent posterior inference across doses and algorithm-like decision rules. These designs (mTPI, mTPI-2, BOIN) are also referred to as model-assisted designs.     

All the  aforementioned   designs are developed for cytotoxic drugs  the toxicity and efficacy of which are assumed to   increase with the dose level. In %the new era of  
modern oncology drug development, novel therapeutics like  kinase inhibitors, monoclonal antibodies, and immunology drugs behave  differently  from traditional cytotoxic agents in that efficacy may plateau or sometimes even decrease as dose level increases \citep{fda2023optimizing}. Consequently, MTD may no longer be optimal for patient care as a lower dose may be as potent but safer.  Instead of finding the MTD, modern  phase I studies are encouraged to identify the optimal biological dose (OBD) that balances the tradeoff between toxicity and efficacy. Ideally, OBD should be no higher than MTD. Apparently, the identification of  the OBD is not achievable by only considering  the dose-limiting toxicity (DLT) data from the phase I studies. Here we aim to propose a model to incorporate pharmacokinetics (PK), pharmacodynamics (PD) information, toxicity and efficacy outcomes of the tested drug in a seamless phase I-II trial for dose optimization. The proposed method, called PEDOOP, is an attempt to respond to the FDA's Project Optimus. Specifically, PEDOOP aims to integrate pharmacological and clinical data to identify an OBD through dose escalation and randomized dose comparison. Even though PK and PD analyses are routinely performed in oncology dose-finding trials, they are not  formally or quantitatively   integrated with clinical data in a dose-finding design. In PEDOOP, we model the time-course PK measurements of each patient and use a summary statistics like the maximum serum concentration or the area under the concentration-time curve (AUC) \citep{piantadosi1996improved,patterson1999novel,whitehead2001easy,whitehead2007bayesian}. 
\citet{ursino2017dose} for integrative modeling. Specifically, the summary statistics of PK is used  as a covariate in a logistic model for the toxicity outcome. We find that this  improves   the estimation of the toxicity probabilities across the doses. % the toxicity probability of doses in our proposed model, where a logit relationship is assumed between the toxicity probability and the logarithm of the AUC. Unlike these previous methods, more detailed PK information, drug concentrations measured at fixed times after administration, is also utilized in our model to produce more reliable estimations.

Furthermore, PEDOOP follows the idea in \citep{su2022semi} and derives a PD summary, which is subsequently linked to efficacy outcomes. %Besides, those methods in the area have some limitations such as the failure to link the PD and the clinical outcomes \citep{su2022semi}. Summary statistics may limit the information we gain from PK and PD profiles. Also, summary statistics which are often calculated by empirical methods or mechanistic models are not precisely and without strength borrow across individuals. Then 
\cite{su2022semi} propose a semi-mechanistic dose-finding (SDF) design based on the modeling of PK and PD profiles and their effects on the DLT probability. A PK-enabled dose-finding model  is developed by the authors to incorporate  patient-level   PK data into phase I. They use drug concentrations to model a latent PD outcome,  which are subsequently linked to DLT outcome.  %and finally linked the DLT probability. 
We modify the SDF design by assuming that the PD is  associated to the efficacy of the new drug, rather than toxicity. %whereas the PK is more related to the toxicity. 
In summary, PEDOOP consists of two modeling components: one for  toxicity  with  PK  as covariate, and the other for efficacy with PD  as  covariate. Both components are derived from the patient-level drug concentrations. %We first modeled the PK profile at the individual level and then obtained the dose-level AUC and PD by integration. Eventually, the dose-level AUC and PD, rather than patient-level ones, are used to model the toxicity and efficacy probabilities. We also compared the simulation results of models that included or excluded the patient-level drug concentration data.

%Another innovative part of this article is that we include the simulation of phase II trials to prove the practicality of our proposed model. As discussed above, more and more phase I studies are targeted at determining the OBD instead of the MTD. However, the advantages of the OBD over other doses are not shown in phase I trials because of the small sample size in phase I. Therefore, phase II trials are included in the simulation to demonstrate the advantages of the OBD selected by the proposed model. In this part, 
For trial design, we follow the idea in the SEARS design \citep{pan2014phase} %. They proposed a phase I/II seamless dose finding model, also known as SEARS, 
that combines phase I dose escalation based on toxicity with phase II randomized  dose comparison based on efficacy.   This idea directly addresses some recommendations in the recent FDA's draft guidance on oncology dose-finding trials. \citep{fda2023optimizing}    %is a very meaningful idea that allows extension from phase I to phase II under one design with no gap in between. 

\begin{figure}[!h] 
  \centering
  \includegraphics[width=\textwidth]{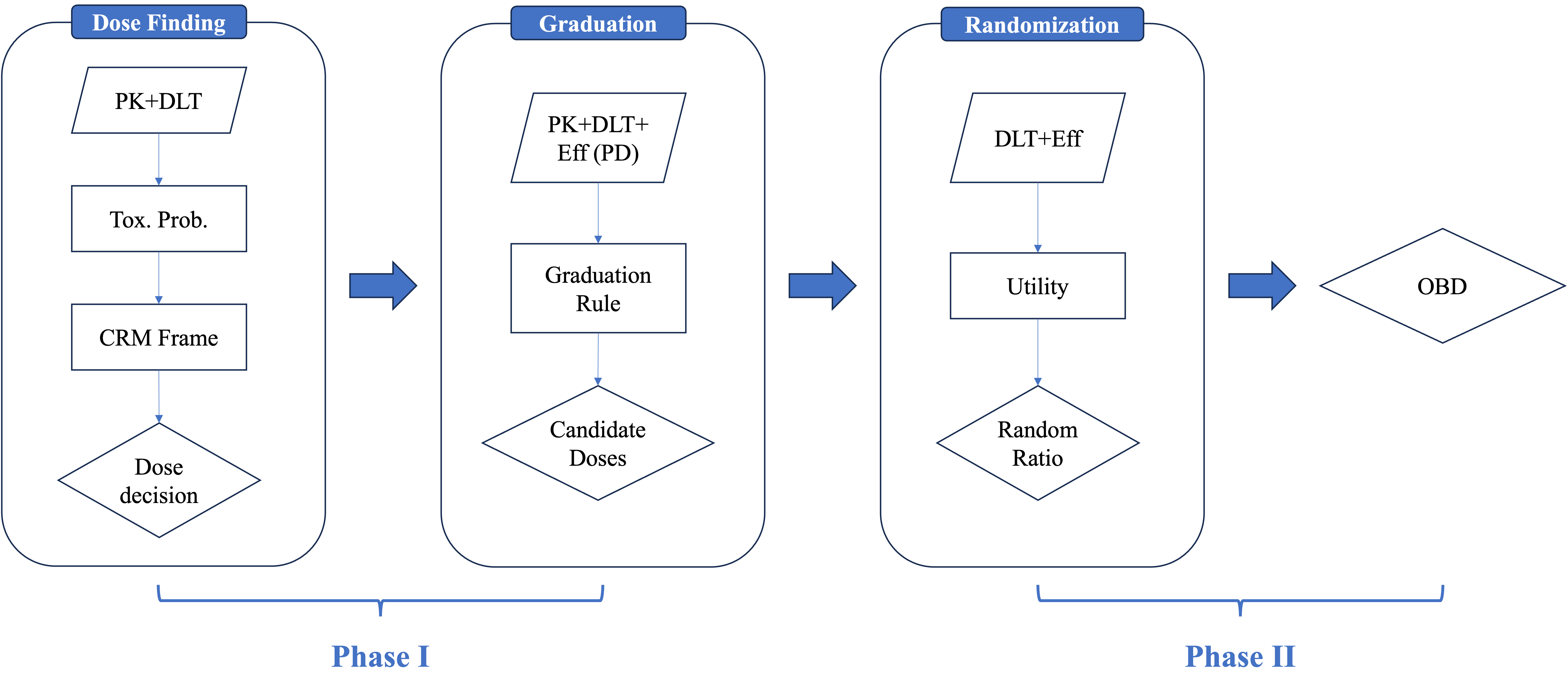}
  \caption{Schema of the PEDOOP design. In Phase I, a PK-enabled design guides dose finding and a graduation rule incorporates PK, DLT, and Efficacy (enhanced by PD) decides which doses are selected for randomized comparison. Phase II involves a randomized comparison of selected doses after which an OBD is selected.}
  \label{fig:schema_PEDOOP}
\end{figure}

A schema of PEDOOP is presented in Figure \ref{fig:schema_PEDOOP}. We firstly calculate the DLT probability based on PK data and DLT outcomes, and conduct dose escalation  using  the CRM frame \citep{o1990continual}. Then, once all efficacy outcomes are collected, graduation rules are used to select doses from all doses in phase I by considering PK data, DLT, and efficacy outcomes. Then,  the adaptive randomization (AR) method is used to adjust the patient allocation ratio among the selected doses from phase I and an optional control arm during the simulated phase II trials. In the end, an OBD will be selected for further study, or the tested drug will be declared non-promising.   We assume the patients in phase I are with different indications and those in phase II are restricted to a specific indication.    

The remainder of the article is organized as follows. Sections \ref{sec:phase1}  and \ref{sec:phase2} describe the phase I and phase II designs of PEDOOP, respectively.  %  Details of the proposed PEDOOP design phase II design are introduced in Section . 
In Section \ref{sec:simluation}, we conduct simulation studies  to  %and through simulation results we 
demonstrate the design performance of PEDOOP. % of the proposed PEDOOP design. 
 We  end the paper with a discussion in  Section \ref{sec:discussion}.  %, the discussion part of the article. 

\section{Phase I Design} \label{sec:phase1}
The proposed PEDOOP design enrolls patients in cohorts, say three patients per cohort. After a cohort of patients is enrolled and treated at a dose, they are followed for a period of time for PK, toxicity, and efficacy outcomes. Specifically, PEDOOP models the DLT outcomes and time-series drug plasma concentration for each patient in phase I, and at the end of phase I, also models efficacy outcomes. %Similar to most existing phase I designs, our proposed design is also cohort-based. After a cohort of patients, say three patients, are enrolled into a phase I trial, we wait until the toxicity observation of this cohort is completed and then make a dose decision based on the data collected for all cohorts. Instead of only considering DLT data, we also include drug plasma concentration data to assist in the estimation of doses' toxicity probabilities. Besides, at the end of the phase I trial where the efficacy evaluation of all patients is finished, we propose to include efficacy outcomes in order to recommend doses for the next phase. 
In other words, three types of information for each patient are collected for PEDOOP in the phase I portion, the drug plasma concentrations over a fixed schedule, the binary toxicity outcome, and the binary efficacy outcome. 
Let $X_{ij}$ denote the drug plasma concentration measured at time $t_j$ for patient $i$, $Y_i$ the binary toxicity outcome, and $Z_i$ the binary efficacy outcome. Here $Y_i = 1$ represents patient $i$ experiences the DLT after  administered by the tested drug, and $Y_i = 0$  not. Similarly, $Z_i = 1$ represents patient $i$ shows an efficacy response (such as objective response), and $Z_i = 0$  not. 
DLT outcomes are recorded  for a period of time after drug administration, such as 21 or 28 days. Drug concentration is typically measured at fixed time points during the first week of drug administration. The efficacy outcome, however, takes much longer to measure, usually after 12 weeks of treatment. %such as 3 months. 
Therefore, we do not model efficacy data for making dose escalation decisions. Instead, efficacy data are used for selecting doses for randomized comparison. %In this case, when we do dose finding in the phase I trial, we temporarily ignore the efficacy outcomes and only rely on toxicity outcomes and drug concentrations because of the lag of efficacy data. However, at the end of the phase I trial, for better dose selection, we recommend including efficacy data.
PEDOOP   links  $X_{ij}$, $Y_i$, and $Z_i$ in a clever way, which will be introduced in %model directly. Here, we adopt two intermediate variables, the AUC and the PD outcome, to connect them as shown in 
Sections \ref{sec:toxic} and \ref{sec:effic}. %, respectively.

\subsection{Drug Concentration}
Let $d_i$ denote the drug dosage administered to patient $i$,  $d_i \in \{1, \cdots,D \}$, where $D$ is the total of tested doses. For simplicity, in this paper, we use a one-compartmental model \citep{shargel1999applied} for drug concentration over time, which assumes that the human body can be treated as a single homogeneous compartment. The model is popular in studies where drugs are administered through the intravenous (IV) injection. % and then assume that the drug concentration is the same everywhere. 
Denote by $c(t)$ the true drug concentration at time $t$. The first-order one-compartment model is %for the IV injection as , 
given by
\begin{equation*}
\frac{d c_i(t)}{d t} = -k_i t \text{ with  an  initial value } c_i (0) = \frac{d_i}{V_i}. 
\end{equation*}
Solving the differential equation, we have
\begin{equation} \label{eq:concentration}
 c_i(t \mid d_i, V_i, k_i) = \frac{d_i}{V_i}e^{-k_i t}. 
\end{equation}
Here, $k_i$ and $V_i$ are the patient-specific elimination rate and volume of distribution, respectively, and $t$ is the time after the drug administration. In practice, $c(t)$ is measured with error and we denote $X_{ij}$ the observed concentration at time $t_j$ for patient $i$. %\eqref{eq:concentration} is derived from the following differential equation, 
%Let $\theta_{PK, i} = (V_i, k_i)$ denote the PK parameters for patient $i$. %$X_{ij}$ denote the observed drug plasma concentration measured at time $t_j$ for patient $i$. 
We assume $\log(X_{ij}) = \log\left(c_i(t_j \mid d_i, V_i, k_i)\right) + \epsilon_{ij},$
where $\epsilon_{ij}$ is the random error. It is common to impose a normal distribution on $\epsilon_{ij}$ with mean $0$ and standard deviation $\sigma$, i.e., $\epsilon_{ij} \sim N(0, \sigma^2)$. Therefore, 
\begin{equation*}
    \log(X_{ij}) \sim N\left(\log\left(c_i(t_j\mid d_i, V_i, k_i)\right), \sigma^2\right).
\end{equation*}
    % which is equivalent to
% $$X_{ij} \sim LogN(\log(c_i(t_j\mid d_i, V_i, k_i)), \sigma^2). $$ 
% where $LogN(\mu,\sigma^2)$ denotes a log-normal distribution and its logarithmic mean and deviation are $\mu$ and $\sigma$, respectively.

 In subsequent discussion, we will link PK with toxicity outcomes at each dose. This requires a summary of drug concentration across patients assigned to the same dose. To this end, we propose a prior distribution for  $V_i$ and $k_i$ so that an average dose-level drug concentration can be computed by integrating them out. Specifically, let  %to compute dose-level parameters. We assume that 
\begin{equation} \label{eq:prior_vk}
 k_i \sim Gamma(\alpha_k, \lambda_k), \quad V_i \sim Gamma(\alpha_V, \lambda_V)
\end{equation}
Here, $Gamma(\alpha, \lambda)$ denotes a gamma distribution with the shape parameter of $\alpha$ and the rate parameter of $\lambda$. The prior distributions describe the variabilities associated with patients, as each individual physiological condition may vary. 
%In order to have a closed-form solution of dose-level parameters, such as \eqref{eq:AUC_d}, we assume that $\alpha_V > 1$ and $\alpha_k >1$. 
By integrating  $k_i$ and $V_i$  over  their prior distributions across patients assigned to the same dose,   we derive the  dose-level  drug concentration at dose $d$ as 
\begin{equation} \label{eq:concentration_d}
	\begin{aligned}
		c(t \mid d) &= \iint c(t \mid d, V, k)g_{V}(V)g_{k}(k)d V d k\\
		 		&= \frac{d\lambda_V}{\alpha_V-1}\left(\frac{\lambda_k}{\lambda_k+t}\right)^{\alpha_k}
	\end{aligned}
\end{equation}
where $g_{V}(V)$ and $g_{k}(k)$ are the probability density functions of the prior distribution of $V$ and $k$ as per \eqref{eq:prior_vk}. We assume $\alpha_V > 1$ to ensure a positive value for $c(t \mid d).$ 

\subsection{Link PK and Toxicity} \label{sec:toxic}

Using \eqref{eq:concentration}, we can easily obtain the AUC for patient $i$ by integrating $c_i(t)$ from 0 to $t_{ref}$, a prespecified time, i.e., 
\begin{equation*}
 AUC(d_i, V_i, k_i; t_{ref}) = \int_0^{t_{\text{ref}}}c_i(t \mid d_i, V_i, k_i)dt = \frac{d_i}{V_i k_i}\left(1-e^{-k_it_{\text{ref}}}\right)
\end{equation*}
By setting $t_{ref}$ to infinity, we obtain %In general, the drug concentration is very small at $t_{\text{ref}}$, close to 0, and it may be wise to set the drug concentration at this point to 0 so that we can trend $t$ toward infinity. In that case, within this segment from 0 to  infinity, the value of 
the expected AUC for patient $i$ as %can be written without saying $t_{\text{ref}}$. So, we have
\begin{equation} \label{eq:auc_i}
AUC_i \equiv AUC(d_i, V_i, k_i; \infty) = \int_0^{\infty}c_i(t_j \mid d_i, V_i, k_i)dt = \frac{d_i}{V_i k_i}.
\end{equation}
Similarly, we obtain the dose-level AUC for dose $d$ by integrating $V_i$ and $k_i$ out of \eqref{eq:auc_i} over their prior distributions \eqref{eq:prior_vk}. %with the integrand \eqref{eq:auc_i}. Or 
Mathematically, this is equivalent to integrating \eqref{eq:concentration_d} from 0 to the infinity. As a result, we derive the dose-level AUC for dose $d$ via %At the dose level, the AUC can be written as:
\begin{equation} \label{eq:AUC_d}
AUC(d \mid \alpha_V, \lambda_V, \alpha_k, \lambda_k) = \int_0^{\infty}c(t \mid d) dt = \frac{d\lambda_V\lambda_k}{(\alpha_V-1)(\alpha_k-1)}.
\end{equation}
We assume $\alpha_V > 1$ and $\alpha_k > 1$ to ensure a positive value for the AUC value. For simplicity, we let $\thetaPK  = (\alpha_V, \lambda_V, \alpha_k, \lambda_k)$ for notation purpose.

Finally,  we assume a linear relationship between the logit of toxicity probability $p_d$ and the logarithm of the AUC, given by  
\begin{equation} \label{eq:p_d}
\logit(p_d) = \logit\left[p(d\mid \thetaPK)\right] = \beta_0 + \beta_1 \times \log\left[AUC(d \mid \thetaPK)\right].
\end{equation}
Denoting by $Y_i$ the binary DLT outcome of  patient $i$,  we assume a binomial distribution 
$$Y_i \sim Binomial\left(p(d_i\mid \thetaPK)\right).$$ 

\subsection{Link PK, PD, and Efficacy} \label{sec:effic}
We now develop a new model that links PK to a latent PD and efficacy. %The second part of the phase I design is the efficacy side which describes the relationship between PK data and efficacy probability. When the clinical outcome is effective, the drug effect intensity may be measured by a biomarker. 
\cite{su2022semi}  modeled a latent composite outcome to describe the relationship between the drug concentration and the pharmacologic effect. We follow their idea and model a latent composite PD outcome as a function of drug concentration at a steady state. We consider the sigmoid Emax model, a classic and widely used model in the PD analysis. Let $r(t \mid d)$ be the drug effect intensity between time $t$ and $t+\Delta t$ when $\Delta t \rightarrow 0$, which is the intensity of drug at time $t$. Assume 
\begin{equation}\label{eq:r_d}
r(t \mid d) = \frac{\emax \times c(t \mid d)^{\gamma}}{\edft^{\gamma} + c(t \mid d)^{\gamma}}, 
\end{equation}
where $\emax$ is the maximum possible drug effect, $\edft$ is the drug concentration that causes 50\% of $\emax$, and $\gamma$ is an operational shape (also called ‘hill’) factor that allows a better data fit \citep{meibohm1997basic, ting2006dose}. Similar to the toxicity model, by integrating $r(t \mid d)$ from 0 to infinity, we derive  the cumulative effect of the drug for dose $d$, $\eta(d)$, given by 
\begin{equation} \label{eq:eta_d}
	\begin{aligned}
		\eta(d)&= \int_0^{\infty} r(t \mid d)dt = \int_0^{\infty}\frac{\emax \left( \frac{d \lambda_V \lambda_k^{\alpha_k}}{\alpha_V - 1}\right)^{\gamma} \left( \frac{1}{\lambda_k+t}\right)^{\gamma\alpha_k} }{\edft^{\gamma}+\left( \frac{d \lambda_V \lambda_k^{\alpha_k}}{\alpha_V - 1}\right)^{\gamma} \left( \frac{1}{\lambda_k+t}\right)^{\gamma\alpha_k}} dt \\
            & = \frac{\emax \left( \frac{d \lambda_V \lambda_k^{\alpha_k}}{\alpha_V - 1}\right)^{\gamma}}{\edft^{\gamma}} \int_0^{\infty}  \frac{1 }{(\lambda_k+t)^{\gamma\alpha_k} + \left( \frac{d \lambda_V \lambda_k^{\alpha_k}}{\alpha_V - 1}\right)^{\gamma} / \edft^{\gamma} } dt \\
		&\xlongequal{x=\lambda_k+t} C(d)\emax \int_{\lambda_k}^{\infty} \frac{1}{x^{\gamma\alpha_k} + C(d)} dx 
	\end{aligned}
\end{equation}
where $C(d) = \left( \frac{d \lambda_V \lambda_k^{\alpha_k}}{\alpha_V - 1}\right)^{\gamma} / \edft^{\gamma}$.
For a given value of  $\gamma * \alpha_k > 1$, there  is a closed form for \eqref{eq:eta_d}.  Hereinafter, we  denote   $\phi = \gamma * \alpha_k$. 
For example, if we let  $\phi = 2$ ,  equation \eqref{eq:eta_d} becomes
\begin{equation*}
\eta(d) = C(d) \emax \times \frac{\arctan\left(\frac{x}{\sqrt{C(d)}}\right)}{\sqrt{C(d)}} \bigg|_{\lambda_k}^{\infty}
=\sqrt{C(d)} \emax  \left[ \frac{\pi}{2} - \arctan\left(\frac{\lambda_k}{\sqrt{C(d)}}\right) \right].
\end{equation*}
 The integration in  \eqref{eq:eta_d} is solvable once $\phi$ is given. For mathematical simplicity,  we assume a categorical prior distribution on $\phi$, $\phi \sim \text{Cat}(4,\boldsymbol{p}),$
where $\phi \in \{2,3,4,5\}$ and $\boldsymbol{p} = (0.25,0.25,0.25,0.25)$.   See Appendix \ref{appendix:prior_phi} for more details. 

%  Considering the formulation in \eqref{eq:concentration_d} and \eqref{eq:r_d}, the product  $\alpha_k * \gamma $ affects  the decaying rate of  the drug effect intensity $r(t\mid d)$ over time. By setting the product of the two to be a constant, the proposed model is limiting the rate of decay on $r(t | d)$. Hereinafter, we set the product value to be 2 for convenience.   

Next we  choose a link function $h(\eta)$ to connect the efficacy probability and the PD drug effect. The link function should satisfy \citep{su2022semi} 
\begin{enumerate}
  \item $h(\eta) \in [0, 1]$
  \item $h(\eta)$ is non-decreasing with respect to $\eta$
  \item $h(0) = 0$ and $\lim_{\eta\to\infty} h(\eta) = 1$. 
\end{enumerate}
The first criterion means that the minimum of the efficacy probability is 0, and the maximum is 1. The second  means that  efficacy is non-decreasing with PD, and the third one means that when the PD effect of the drug reaches infinity or zero, the drug must be efficacious with probability one or zero, respectively. %, and if the drug does not act on the biomarker, the drug must be ineffective. 
An example of the link function is $h(\eta) = 1-\exp(-\eta)$ in \cite{su2022semi}, which is the one chosen here for PEDOOP. Therefore, %also that we choose in the proposed model. So, we have
\begin{equation}\label{eq:q_d}
q_d \equiv q\left(d \mid \thetaPK, \emax, \edft, \phi\right) = 1 - \exp(-\eta(d)). 
\end{equation}
Here, $q_d$ is the efficacy probability of dose $d$. Finally, denoting by $Z_i$ the binary efficacy outcome of patient $i$, we assume 
$$Z_i \sim Binomial\left(q(d_i \mid \thetaPK, \emax, \edft, \phi)\right).$$

\subsection{Posterior Inference}
Let $\bm{\theta}$ denote all the dose-level parameters in the model, i.e., $\bm{\theta} = (\alpha_V, \lambda_V, \alpha_k, \lambda_k, \sigma^2, \beta_0, \beta_1, ED_{max}, \edft, \phi)$. Denote by $\mathcal{D}_{\text{I}} = \{(X_{ij}, Y_i, Z_i, d_i)\}$ all the available data from Phase I. The joint posterior distribution can be written as 
\begin{equation} \label{eq:posterior_inf}
  \begin{aligned}
    \pi(\bm{\theta}\mid \mathcal{D}_{\text{I}}) \propto &\prod_i\prod_j \frac{1}{\sigma \sqrt{2\pi}}\exp \left(-\frac{(\log(X_{ij})-\log\left(c_i(t_j \mid d_i, V_i, k_i)\right)^2}{2\sigma^2} \right) \times \prod_i f(V_i, k_i \mid \alpha_V, \lambda_V, \alpha_k, \lambda_k) \\
    & \times \prod_d q_d^{z_d}(1 - q_d)^{n_d-z_d}\\
    & \times \prod_d p_d^{y_d}(1 - p_d)^{n_d-y_d}\\
    & \times g(\bm{\theta}), 
  \end{aligned}
\end{equation}
where $n_d = \sum_i I(d_i = d)$, $y_d = \sum_i I(d_i = d, Y_i = 1)$, and $z_d = \sum_i I(d_i = d, Z_i = 1)$, which represent the number of patients, patients experiencing DLT, and patients with efficacy response at dose $d$, separately. 
The factors in the four lines of \eqref{eq:posterior_inf} correspond to the likelihood function for the PK data and prior (the first line), efficacy response data (the second line), toxicity response data (the third line), and prior for the dose-level parameters (the last line). Here, $g(\bm{\theta})$ denotes the prior distribution of $\bm{\theta}$, which will be elaborated in Section \ref{sec:simluation}.

We use Markov chain Monte Carlo (MCMC) simulation to generate samples from the posterior distributions of the unknown parameters, with which statistical inference is conducted.  The computation is implemented using R/JAGS.  

\subsection{Design Algorithm} \label{sec:phase-I-algo}
Due to the model-based nature of PEDOOP, we follow the principle of %The design algorithm in the proposed design is similar to 
the CRM design \citep{o1990continual}. Specifically, PEDOOP  % which estimate doses' toxicity probabilities as the posterior mean and make dose decisions according to the distance of that from the target toxicity rate $p_T$. The details are as follows:
% Let $\textbf{d} = (d_1, d_2, \cdots, d_n)$ denotes the n doses in phase I study and $n_{max}$ denotes the maximum sample size the the trial. 
allocates the first cohort of patients at the starting dose and continuously allocates subsequent cohorts to a dose that is deemed more beneficial to patients. Similar to the CRM design, % 
%the available data $\mathcal{D}_{\text{I}}$ is updated once a cohort of patients  completes toxicity observation. 
 let $\hat{p}_d$ denotes the  posterior mean of the toxicity probability of dose $d$ given the current trial data $\mathcal{D}_{\text{I}}$. PEDOOP finds the dose $d'$ for the next cohort given by
$$
d' = \argmin_{d} |\hat{p}_d - p_T|.
$$
However, if there is no DLT at the current dose and doses below the current dose, the next cohort is always treated at the dose one  level above the current dose to speed up exploration. That is, $d' = d +1$ if $\sum_i^{d} y_i = 0$, where $d$ represents the current dose in the dose escalation. Furthermore, for safety,  if $d^* = \mathop{\arg\min}\limits_{d} |\hat{p}_d - p_T|$, $d^* > d+1$, and $n_{d^* - 1} = 0$, then $d' = d +1$. This is essentially no-skip in dose escalation, a common practice by the CRM design. 

During the trial, a safety rule is added to avoid overdosing or terminate the trial if there are no safety doses. If dose $d$ satisfies
$$
\Pr( p_d > \pi_S \mid \mathcal{D}_{\text{I}}) < s^*, 
$$
then dose $d$ is considered safe. Here, $\pi_S$ and $s^*$ are two prespecified parameters, and we can easily let $\pi_S = p_T$ and $s^* = 0.95$.  

% \par The allocation scheme of the design is summarized as \ref{alg:cap}:
% \begin{algorithm}
% \caption{phase I design }\label{alg:cap}
% \begin{algorithmic}
% \Require{Allocate the first cohort of patients at the starting dose. }
% \While{the number of cohort $>$ 0}{
% \\ Using the model to calculate the toxicity probability $\hat{p}_d$ for each dose and get the optimal dose $d^*$. \\
%   Allocate the next cohort at $d^*$ with untried doses. \\
%   We will always escalate unless the total number of DLT is large than 0. }
%   \If{There is no safe dose}
%   \\ terminate the trial
%   \EndIf 
% \EndWhile
% \end{algorithmic}
% \end{algorithm}

\subsection{Graduation Rule} \label{sec:phase-I-grad}
At the end of phase I, %Once the accrual of the phase I trial is completed and the last cohort of patients finishes follow up completed efficacy evaluation, several 
 %more than one dose will be recommended to the next phase.
% Generally speaking, 
if a dose exhibits promising efficacy and low toxicity, it will graduate to  phase II. We follow the SEARS design \citep{pan2014phase} and implement the following graduation rule: if dose $d$ meets the two criteria in \eqref{eq:gra_rule}, it will graduate  to the next phase and be compared in a randomized fashion.
\begin{equation} \label{eq:gra_rule}
  \Pr(p_d < \pi_T\mid \mathcal{D}_{\text{I}}) > p^* \text{ and } \Pr(q_d > \pi_E\mid \mathcal{D}_{\text{I}}) > q^*,
\end{equation}
where $p_d$ and $q_d$ denote the toxicity and efficacy probabilities of dose $d$, respectively. 
$\pi_T$, $\pi_E$, $p^*$ and $q^*$ are four physician-specified values. For example, $\pi_T$ can be set to the target toxicity probability in phase I and $\pi_E$ can be the historical response rate of the standard treatment.  If multiple doses satisfy the rules in \eqref{eq:gra_rule}, all of them are allowed to graduate to phase II depending on the resources.   And if no dose satisfies the rules in \eqref{eq:gra_rule}, the trial stops.

\section{Phase II Design}\label{sec:phase2}

\subsection{Utility} \label{sec:utility}

Assume doses $\{r \mid r= 1, \cdots, R\}$ graduate from the phase I as per the graduation rule \eqref{eq:gra_rule}, which will be regarded as separate arms in phase II. Besides that, we assume that a control arm is included which could be the standard of care. Let $r = 0$ denote the control arm. The patient population in phase II is now restricted to a specific indication and enrolled patients are adaptively randomized to arms $\{r \mid r= 1, \cdots, R\}$ and the control arm. We use a joint  efficacy and toxicity outcome as the primary endpoint for the phase II trial. %The objective of the phase II design is to compare the efficacy of different doses relative to the control. %verify whether one of $\{r\mid r= 1, \cdots, R\}$ is the most promising for further study, or none of them is not promising compared to the control arm.
There are four possible bivariate binary outcomes for a patient, (No toxicity, No efficacy), (No toxicity, Efficacy), (Toxicity, No efficacy), and (Toxicity, Efficacy). We use a utility function to numerically score %It might be more complicate to consider both toxicity and efficacy outcomes. Therefore, we score 
the four outcomes  in the table below. %in advance, that is, assigning utility, to better evaluate these arms, $\{r\mid r= 0,1, \cdots, R\}$ in section \ref{sec:utility}. Then, a quasi-binomial distribution constructed on the utility to help conduct the adaptive randomization (AR) in section \ref{sec:patient_allocation}.
%The following $2\times 2$ table shows the notations for the utilities of the four types. 
The four  utility values $\{s_1, s_2, s_3, s_4\}$ are used to describe the relative benefits of the treatment to a patient. We assume each $s_i \in [0, 1]$ without loss of generality. Apparently, $s_1$ should be the largest since it is associated with the best outcome (No toxicity, Efficacy) for a patient. We set $s_1 = 1$. Conversely, we let $s_4 = 0.$ The other two values, $s_2$ and $s_3$, are between 0 and 1 and are elicited from clinicians.  Their values are contingent upon the specific circumstances and conditions encountered in real clinical trials.   Specifically, for very aggressive cancers, $s_2$ could be larger than $s_3$ to reflect the desire for efficacy, while for less aggressive cancers, one may prefer a smaller $s_2$ than $s_3$.  In the proposed design, $s_2$ and $s_3$ are set to 0.6 and 0.4 temporarily.   %\note{  Should we cite Ying Yuan's paper and explain how $s_2$ and $s_3$ are elicited?     It's a different Yuan. The quasi-binomial likelihood was proposed to accommodate toxicity grades in \cite{yuan2007continual}, not for the combination of toxicity and efficacy outcomes. But the concepts of utility of joint outcomes was also adopted by Ying Yuan in U-BOIN. However, I feel there's no need to add how $s_2$ and $s_3$ are elicited. Because,  as described in the text, these values should be derived from clinicians' experience. In the field of statistics, there is no best choice.  }  

\begin{table}[htbp]
\centering
\caption{Utility of the joint efficacy toxicity outcomes. Each $0 \le s_i \le 1$ value represents the relative benefit of the outcome for a patient.}
\begin{tabular}{|c|p{1cm}|p{1cm}|}
\hline
\diagbox{Efficacy}{Toxicity} & No & Yes \\ \hline
Yes   & $s_1$  & $s_2$  \\ \hline
No    & $s_3$  & $s_4$  \\ \hline
\end{tabular}
\end{table}

Let $p_{r1}$, $p_{r2}$, $p_{r3}$, and $p_{r4}$ denote the probabilities of (No toxicity, Efficacy), (Toxicity, Efficacy), (No toxicity, No efficacy), and (Toxicity, No efficacy)  for   arm $r$, respectively. Note that $(p_{r1}+p_{r2})$ also represents the toxicity probability and $(p_{r2}+p_{r4})$ the efficacy probability. Given the probabilities $(p_{r1}, p_{r2}, p_{r3}, p_{r4})$ and the utilities $(s_1, s_2, s_3, s_4)$, we can calculate the expected utility of arm $r$ by
$$
u_r = \sum_{i=1}^4 s_i\times p_{ri}.
$$

\subsection{Patient Allocation} \label{sec:patient_allocation}
%We propose to model $u_r$  based on the quasi-likelihood theory. If we set $s_1 = 1$ and $s_4 = 0$. Then $r_d \in [0, 1]$ can be viewed as a weighted average of $(p_{r1}, p_{r2}, p_{r3}, p_{r4})$ and also as a utility probability.
Let $(y_{r1}, y_{r2}, y_{r3}, y_{r4})$ denote the numbers of patients at dose $d$ having outcomes (No toxicity, Efficacy), (Toxicity, Efficacy), (No toxicity, No efficacy), (Toxicity, No efficacy), respectively. Then the total utility of patients at arm $r$ can be calculated by $ S_r = \sum_{i = 1}^4 s_i\times y_{ri} $.
And $S_r$ can be interpreted as the number of ``events" observed among $n_d$ patients treated at arm $r$ given the event probability $u_r$. Following \citet{yuan2007continual} and \citet{lin2020boin12},  given $\{s_i; \; i=1,\ldots, 4\}$, we model %So, we can model $u(d)$ using the binomial distribution with 
the ``quasi-binomial" data $(S_r, n_r)$ with a quasi-binomial likelihood of the observe data $(S_r, n_r)$ is 
\begin{equation*}
   L((S_r, n_r)\mid u_r) \propto u_r^{S_r} (1-u_r)^{n_r-S_r}
\end{equation*}
Under the Bayesian framework, we assign  $u_r$  %$u(d)$ 
a Beta prior, i.e.
$ u_r  \sim Beta(\alpha, \beta)$,
where $\alpha$ and $\beta$ are the prespecified hyperparameters. By default, we set $\alpha = \beta = 1$, i.e., a weakly informative prior distribution for $u_r$. The posterior distribution of $u_r$ arises as 
$$
u_r \mid (S_r, n_r) \sim Beta(\alpha+S_r, \beta+n_r-S_r).
$$
  Note that one could easily incorporate efficacy and toxicity data from phase I into the modeling. Denote the corresponding phase I quasi-binomial data as $(S_{r1}, n_{r1})$. Then the posterior distribution of $u_r$ is $Beta(\alpha+S_r+S_{r1}, \beta+n_r+n_{r1}-S_r-S_{r1}).$   
% The proposed quasi-binomial approach has 2 important advantages. First, it greatly simplifies the posterior calculation of $u(d)$. Second, it is highly scalable in the sense that the model and method are invariant to the dimension of the endpoints. 

%AR aims to assign more patients to more effective arms. 
 Using the posterior distributions, we  consider Bayesian adaptive randomization (BAR) procedures to continuously update the randomization probability for  arm $r$.   %according to the posterior distributions of outcome parameters. %observed response data. 
 For example, a popular BAR approach  randomizes   patients to dose arm $r$ with a probability proportional to the posterior probability that the efficacy probability of arm $r$, $q_r$, is larger than a prespecified threshold $\pi_E$ given the observed data, i.e., $\Pr(q_r > \pi_E \mid \mathcal{D}_{\text{II}})$. Here, $\mathcal{D}_{\text{II}}$ represents the observed data in phase II, and $\mathcal{D}_{\text{II}} = \{(y_{r1}, y_{r2}, y_{r3}, y_{r4}, n_r), \;  r = 0,1, \cdots, R\}$. However, \cite{huang2007parallel} pointed out this approach does not reflect the relative desirability of each arm. Instead, they propose to randomize patients based on  %$\Pr(q_r > \pi_E \mid \mathcal{D}_{\text{II}})$ to 
$\Pr(q_r > \max\{q_{r'}; r'\neq r\} \mid  \mathcal{D}_{\text{II}})$ which can result in more patients treated in arms with relatively high efficacy rates.

We apply a BAR scheme similar to  \cite{huang2007parallel}. To take advantage of the quasi binomial utility, we replace the efficacy probability in their formula with the utility above. Thus, patients are randomized to arm $r$ with a probability proportional to
\begin{equation}\label{eq:prob_maxu}
    \xi_r \equiv \Pr(u_r > \max\{u_{r'};  r'\neq r\}\mid \mathcal{D}_{\text{II}})
\end{equation}

\subsection{Trial Design} \label{sec:trial_design}
We propose to update the randomization probabilities $\xi_r$ after a cohort of $n$ patients' outcomes are obtained, and the trial continues until the maximum sample size $N$ is reached.
The phase II part of the proposed PEDOOP design is as follows.
\begin{enumerate}
    \item {\bf Run in:} The first $n$ patients as a cohort are randomized equally to  the  $(R+1)$ arms.
    \item Update the randomization probabilities $\xi_r$'s in \eqref{eq:prob_maxu}. %according to the posterior quasi-binomial distribution of $u_r$, shown in $\S$\ref{sec:patient_allocation}.
    \item The next cohort of $n$ patients are randomized to arms $\{ 0, 1, \cdots, R\}$ with $\xi_r.$ %the updated randomization probabilities.
    \item Repeat steps 2 and 3 until the maximum sample size $N$ of phase II is reached.
    \item Upon conclusion of the trial, recommend the arm with the highest posterior probability \eqref{eq:prob_maxu} to be further studied. If the control arm has the highest utility, no arm is recommended.
\end{enumerate}
 In practice, efficacy outcomes are usually observed much later than toxicity outcomes. In order to apply the above algorithm one should wait until the efficacy outcomes from the previous cohort are observed before updating the randomization probabilities, which would likely increase the duration of the trial depending on how fast patient accrual is. On the other hand, the proposed adaptive randomization  can potentially increase the chance a patient is assigned to a more effective and safer arm. For a practical trial, the cohort size should be selected based on the  observation time of efficacy and toxicity outcomes, enrollment speed, the number of arms in the trial, among other criteria. One may find a proper cohort size $n$ by balancing the tradeoff among these criteria.

BAR procedures often employ practical steps to avoid undesirable performance. See \cite{thall2007practical} and more recently \cite{robertson2023response} for comprehensive reviews. Since the methodology of BAR is not the focus of this work, we do not go into detail on how BAR could be improved with recommendations in the cited works. In practice, we recommend applying the additional rules in the two papers in order to further improve the performance of the BAR algorithm above.    

\subsection{Arm Selection} \label{sec:arm_sel}
At the end of phase II, %Once the last patient is completed efficacy evaluation, 
several doses among $\{r\mid r= 1, \cdots, R\}$ will be recommended to the next phase.
The SEARS design \citep{pan2014phase} implemented the following selection rule: if arm $r$ meets the two criteria \eqref{eq:sel_rule},  it  %arm $r$ 
will be recommended to the next phase. Let 
\begin{equation} \label{eq:sel_rule}
  C = \left\{r: \Pr(p_r < \pi_T\mid \mathcal{D}_{\text{II}}) > p^{**} \text{ and } \Pr(q_r > \pi_E\mid \mathcal{D}_{\text{II}}) > q^{**}\right\},
\end{equation}
where $p_r$ and $q_r$ denote the toxicity and efficacy probabilities of arm $r$. And the posterior distributions of $p_r$ and $q_r$ are both based on the beta prior, $Beta(1,1)$. $p^{**}$ and $q^{**}$ are two physician-specified values.  Criteria \eqref{eq:sel_rule} will potentially lead to selecting more than one dose, and we call these ``admissible doses".  

%However, \eqref{eq:sel_rule} may result in multiple arms selected from the trial, which is not desirable in practice. Therefore, in  PEDOOP  %our proposed design, 
We also propose to select the  dose   with the highest  utility based on  posterior probability \eqref{eq:prob_maxu} among  those    meeting the criteria \eqref{eq:sel_rule}. That is, the dose $r^*$ is selected 
where
\begin{equation*}%\label{eq:sel_dose}
r^* = \argmax_{r \in C} \xi_r = \argmax_{r \in C} \Pr(u_r > \max\{u_{r'};  r'\neq r\}\mid \mathcal{D}_{\text{II}}).
\end{equation*} 
 Dose $r^*$ is the estimated OBD.

% The Figure \ref{fig:whole_algorithm} is the algorithm flow chart of Phase I and Phase II. 
% \begin{figure}[!h] \label{fig:whole_algorithm}
%   \centering
%   \includegraphics[width=0.80\textwidth]{3.png}
%   \caption{Caption}
% \end{figure}

\section{Simulation Study}\label{sec:simluation}
\subsection{Prior Distribution} 
In our simulation studies, we adopt the following prior distributions for  model parameters. 
\begin{equation} \label{params:PEDOOP}
    \begin{aligned}
    &\alpha_V - 1 \sim Gamma(4,1), \quad \lambda_V \sim Gamma(1,1), \\
    &\alpha_k - 1 \sim Gamma(3,1), \quad \lambda_k \sim Gamma(1,1), \\
    &\sigma \sim Gamma(3,3),\\
    &\beta_0 \sim N(-3,10^2), \quad \beta_1 \sim LogN(-1,2),\\
    &\emax \sim LogN(-1,0.5), \quad \ \edft \sim Gamma(20,0.5), \quad \text{and} \quad \phi \sim \text{Cat}(4,\boldsymbol{p}),
    \end{aligned}
\end{equation}
where $\phi \in \{2,3,4,5\}$ and $\boldsymbol{p} = (0.25,0.25,0.25,0.25)$.
Here, the shape parameters $\alpha_V$ and $\alpha_k$ follow $1+Gamma(4,1)$ and $1+Gamma(3,1)$ distributions. This assumption ensures that $\alpha_V > 1$ and $\alpha_k > 1$, which are necessary conditions for a closed-form solution of dose-level parameters, such as the dose-level AUC in \eqref{eq:AUC_d}. 

\begin{figure}[!htbp]
\centering
\begin{subfigure}{.5\textwidth}
  \centering
  \includegraphics[width=\linewidth]{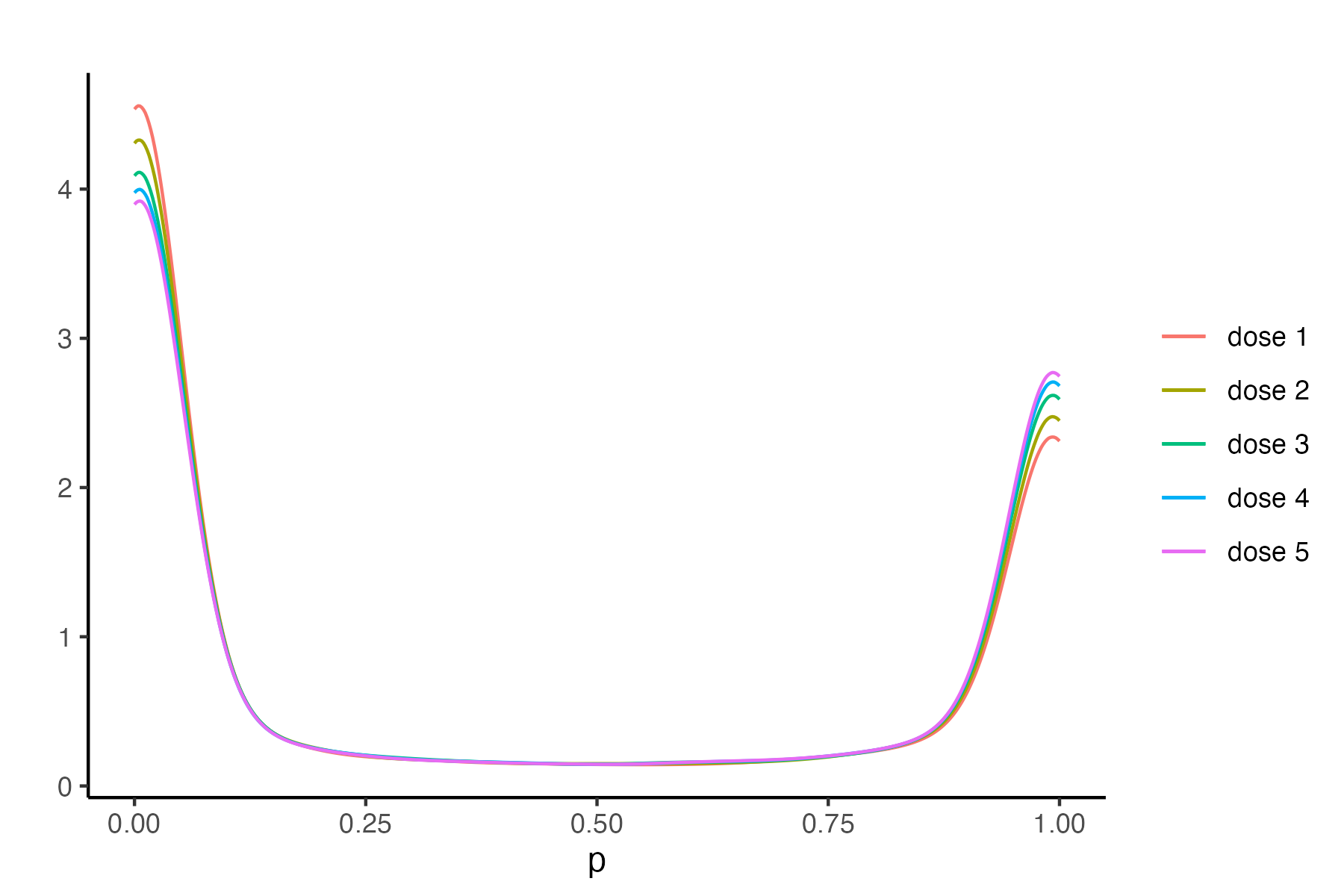}
  \caption{$p_d$ }
\end{subfigure}%
\begin{subfigure}{.5\textwidth}
  \centering
  \includegraphics[width=\linewidth]{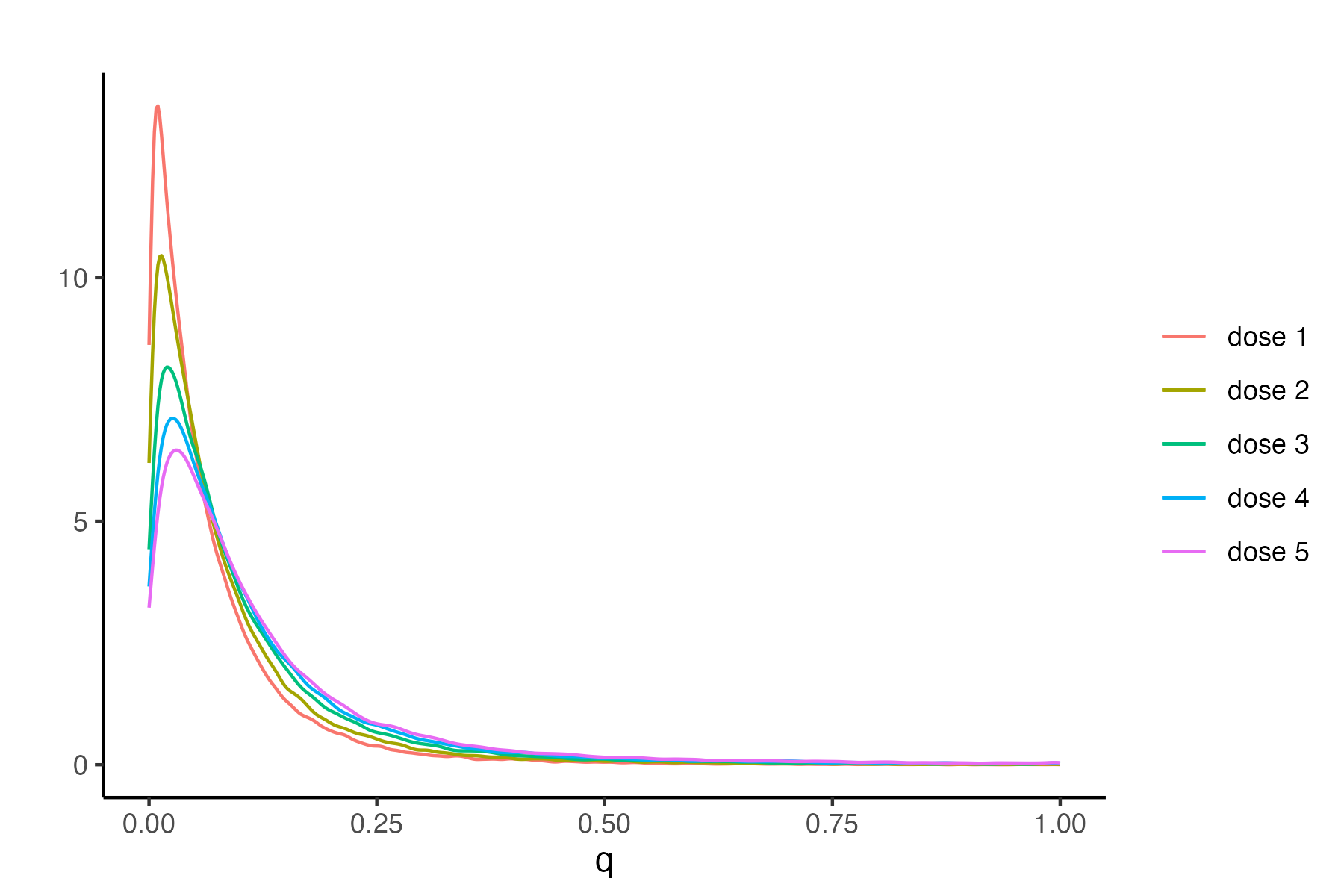}
  \caption{$q_d$}
\end{subfigure}
\caption{Density of prior distributions of $p_d$ and $q_d$, $d = 1, \cdots ,D$, given hyperparameters in \eqref{params:PEDOOP}.}
\label{fig:prior_pred_p}
\end{figure}

The hyperparameters  in %\eqref{params:LE} %and 
\eqref{params:PEDOOP}  are determined so that the corresponding priors of $p_d$ and $q_d$ are vague distributions. Specifically, with the priors in \eqref{params:PEDOOP}, the corresponding prior distribution of $p_d$ can be numerically determined according to equations \eqref{eq:AUC_d} and \eqref{eq:p_d}, and $q_d$ according to \eqref{eq:q_d}. %approaches 0.5 while its density is concentrated around 0 and 1. 
We plot the corresponding prior densities of  $p_d$ and $q_d$, $d = 1, \cdots, D$ in Figure \ref{fig:prior_pred_p}.  The plots show that the priors of $p_d$'s are of small density values in most of the parameter space.

% \begin{figure}[!htbp]
% \centering
% \begin{subfigure}{.5\textwidth}
%   \centering
%   \includegraphics[width=\linewidth]{LE_q_prior.png}
%   \caption{Within DOOP  given hyperparameters in \eqref{params:LE}}
% \end{subfigure}%
% \begin{subfigure}{.5\textwidth}
%   \centering
%   \includegraphics[width=\linewidth]{PEDOOP_q_prior.png}
%   \caption{Within PEDOOP given hyperparameters in \eqref{params:PEDOOP}}
% \end{subfigure}
% \caption{Density of prior  distributions of $q_d, d = 1, \cdots ,D$}
% \label{fig:prior_pred_q}
% \end{figure}

\subsection{Simulation 1} \label{sec:sim1}
 \subsubsection{Simulation Setup}  %\label{sec:sim1_setup}
To  assess the impact of integration of PK data like the time-course  % the strength of incorporating the data of 
drug concentration, we perform 1,000  simulated trials  % simulation studies 
to compare the proposed PEDOOP  design and a simpler model, called ``DOOP'', meaning PEDOOP without P (PK) or E (enabled). Specifically, DOOP adopts the same designs as PEDOOP described in Sections \ref{sec:phase-I-algo}, \ref{sec:phase-I-grad}, and \ref{sec:phase2} but does not incorporate PK data. Instead, DOOP simply    uses   a logit regression model \eqref{eq:simple_p} on toxicity  and an Emax model \eqref{eq:simple_q} on efficacy   without integrating PK data, i.e.,   
\begin{equation}\label{eq:simple_p}
    \logit(\tilde{p}_d) = \tilde{\beta_0} + \tilde{\beta_1} \times \log(d).
\end{equation}
 For DOOP, we adopt similar priors as in PEDOOP.   
\begin{equation}\label{eq:simple_q}
    \logit(\tilde{q}_d) = 1 - \exp(-\tilde{\eta}(d)), \quad \text{where} \quad \tilde{\eta}(d) = \frac{\tilde{E}_{\text{max}} \times d^{\tilde{\gamma}}}{\widetilde{ED}_{50}^{\tilde{\gamma}} + d^{\tilde{\gamma}}}.
\end{equation}
\begin{equation} \label{params:LE}
    \begin{aligned}
    &\tilde{\beta}_0 \sim N(-3,10^2), \quad \tilde{\beta}_1 \sim LogN(-1,2),\\
    &\tilde{E}_{\text{max}} \sim LogN(-1,0.5), \quad \widetilde{ED}_{50} \sim Gamma(10,0.1), \quad \text{and} \quad \tilde{\gamma} \sim Gamma(0.1,0.1).
    \end{aligned}
\end{equation}
The joint posterior distribution of $\tilde{\bm{\theta}} = \{\tilde{\beta_0}, \tilde{\beta_1},  \tilde{E}_{\text{max}} , \widetilde{ED}_{50}, \tilde{\gamma}\}$ can  be  written as
$$\pi(\tilde{\bm{\theta}} \mid \mathcal{D}_{\text{I}}) \propto \left\{\prod_d \tilde{p}_d^{y_d}(1 - \tilde{p}_d)^{n_d-y_d} \right\} \times \left\{\prod_d\tilde{q}_d^{z_d}(1 - \tilde{q}_d)^{n_d-z_d} \right\} \times g(\tilde{\bm{\theta}}), $$
where $g(\tilde{\bm{\theta}})$ denotes the prior distribution of $\tilde{\bm{\theta}}$.

For simplicity, in this section, we only compare the phase I simulation results of PEDOOP and DOOP  since the phase II portion of both designs are identical.   The target toxicity rate is set to $p_T = 0.3$. We assume that there are five dose levels, i.e., $d \in (15, 30, 60, 90, 120)$, and drug concentration are measured for each patient six times, at 1, 3, 5, 7, 12, and 24 hours after administration.
A maximum of 30 patients with three patients as  a cohort is to be enrolled in phase I.
For both designs, $\pi_T = 0.3$, $p^* = 0.6$, $q^* = 0.6$, and $\pi_E = 0.2$ in the graduation rule \eqref{eq:gra_rule}.

Here we generate the patient-specific elimination rate and volume of distribution $k$ and $V$ for 30 individuals randomly through gamma distributions,  and  the generated $V$ and $k$ values are subsequently used to generate drug concentration data $X$ using a normal distribution. For example, for patient $i$, the corresponding values are generated via 
\begin{equation} \label{eq:true_AUC1}
V_i \sim Gamma(4,1), \quad k_i \sim Gamma(3,1), \quad \log(X_{ij}) \mid V_i, k_i  \sim N\left(\log\left(\frac{d_i}{V_i}\right) - k_i t_j,1\right),
\end{equation}
where $d_i \in \{7, 15, 30, 60, 120, 150\}$ and $t_j \in \{1, 3, 5, 7, 12, 24\}$.  
 We set $\emax = 1$, $\edft = 100$, and $\phi = 2$ for all the scenarios. Given the true parameter values, the cumulative effect of the test drug is generated via \eqref{eq:eta_d}, which is substituted into \eqref{eq:q_d} to generate the efficacy probabilities for the five dose levels. This leads to true efficacy probabilities (0.12, 0.18, 0.27, 0.33, 0.37) for all the scenarios. %That indicates  only the third dose or higher exhibits a therapeutic effect exceeding $\pi_E$.
% All the scenarios have the same efficacy probabilities of the five dose levels, (0.12, 0.18, 0.27, 0.33, 0.37). This means only the third or higher dose exhibits a therapeutic effect greater than $\pi_E$. 
To generate true toxicity probabilities, we first compute the dose level AUC via \eqref{eq:AUC_d} and then assign different values of  $(\beta_0,\beta_1)$ scenarios 1 through 4 in \eqref{eq:p_d} as $\{(-2.5, 1), (-3, 0.95), (-4.5, 1.35), (-2.5, 0.5)\}.$ This leads to the true toxicity probabilities in  Table \ref{tab:sim1}. 

\begin{figure}[!htbp]
  \centering
  \includegraphics[width=0.8\textwidth]{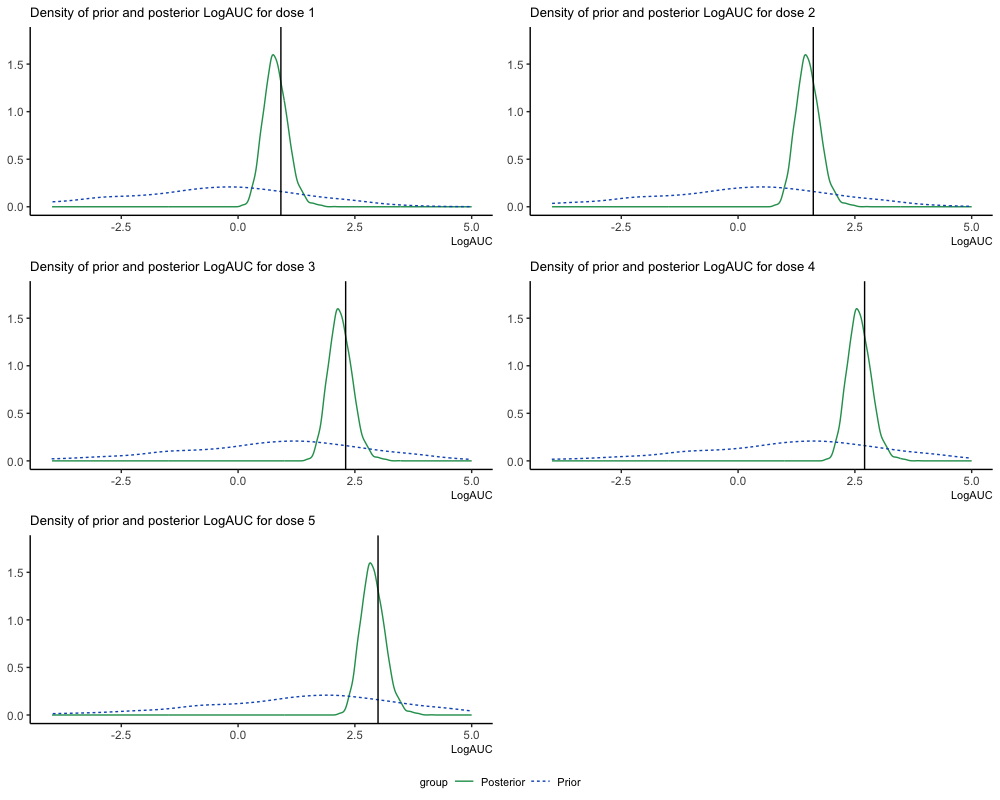}
  \caption{Densities of the prior and posterior distributions of the logarithm of AUC for five doses in a simulated trial. The green solid line represents the posterior distribution, while the blue dashed line represents the prior distribution. And the vertical line denotes the logarithm of the true AUC. }
  \label{fig:auc}
\end{figure}

Figure \ref{fig:auc}  presents   the density plots of the logarithm of AUC for the five doses in a simulated trial. The %density of 
logarithm AUC is numerically  calculated  according to \eqref{eq:AUC_d}. It shows that the posterior distribution is able to shrink the prior towards the truth.

In this simulation study, we set up a total of four distinct scenarios. To create these scenarios, we held the true values of $\bm{\theta}$ constant, thereby determining the toxicity and efficacy probabilities for each one. 

\subsubsection{Simulation Results} 
The operating characteristics of the PEDOOP and DOOP models are summarized in Table \ref{tab:sim1}. 
%In this section, we initiate 
 We compare  %a comparison between 
PEDOOP and DOOP based on two key factors,  patient allocation and  dose selection by the models. We begin by presenting the selection frequency of doses at the end of phase I according to the graduation rule \eqref{eq:gra_rule}.  We refer to these doses the ``admissible doses".  It's worth noting that the graduation rule may lead to the selection of multiple doses at the end of phase I. This feature enables phase II to employ adaptive randomization for a further exploration of these doses. Also, we present the  selection frequency of  
% ``Sel \% with U",  which selects 
doses  based on the graduation rule \eqref{eq:gra_rule} and the highest posterior probability \eqref{eq:prob_maxu}, where  no more than  one dose is recommended to phase II in each simulated trial.  The dose with the highest posterior probability among  the admissible doses is considered  the OBD.

 In general, PEDOOP and DOOP allocate comparable numbers of patients to  the OBDs  in the four scenarios. In view of OBD selection frequency, PEDOOP is more desirable than DOOP. 

In scenario 1, there are no desirable doses since the first two doses of the tested drug have lower efficacy probabilities than $\pi_E$, while the last three have higher toxicity probabilities than $\pi_T$. PEDOOP  exhibits a lower frequency of selecting either dose than DOOP.  % selects these doses less frequently, both in terms of admissible doses and OBD.
 In scenarios 2, 3 and 4, , doses 3, 4, and 5 are the true OBD respectively.  PEDOOP presents a higher frequency of selecting the true OBD in all three scenarios.  %always selects the optimal dose with a higher proportion both in both admissible doses and OBD.

We also investigate the estimation of $\phi$, and present the posterior estimates of $\phi$ in Table \ref{tab:pos_phi}. Overall, the estimates are reasonable, close to its true value of 2.

\begin{table}[!htbp]
  \centering
  \renewcommand{\arraystretch}{0.7}
  \caption{Phase I simulation results of  Simulation 1.  Here, the bold font indicates the true OBD in each scenario, which has a  toxicity rate  closest to  $p_T = 0.3$ and an efficacy rate  larger than $\pi_E = 0.2$. And ``Admissible Doses" and ``OBD" represent the selection frequency as admissible doses and OBD in percentage, respectively.}
    \begin{tabular}{c|c|llllll}
          & \multicolumn{2}{c}{Dose} & 1     & 2     & 3     & 4     & 5 \\
          & \multicolumn{2}{c}{True Eff. Prob.} & 0.12  & 0.18  & 0.27  & 0.33  & 0.37 \\
    \hline
    \multirow{8}[2]{*}{Sc 1} & \multicolumn{2}{c}{True Tox. Prob.} & 0.17  & 0.29  & 0.45  & 0.55  & 0.62 \\
          & \multicolumn{2}{c}{Utility} & 0.404 & 0.392 & 0.382 & 0.378 & 0.374 \\
          & \multirow{3}[0]{*}{PEDOOP} & \# of Patients & 10.494 & 11.163 & 5.79  & 1.587 & 0.627 \\
          &       & Admissible Doses  & 1.5   & 6.6   & 3.5   & 0.1   & 0.2 \\
          &       & OBD  & 1     & 5.6   & 2.8   & 0     & 0.1 \\
          & \multirow{3}[1]{*}{DOOP} & \# of Patients & 10.032 & 11.343 & 6.057 & 1.608 & 0.573 \\
          &       & Admissible Doses  & 22.5  & 18.3  & 4.2   & 0.5   & 0.4 \\
          &       & OBD  & 14.3  & 9.8   & 2.1   & 0.2   & 0.1 \\
    \hline
    \multirow{8}[2]{*}{Sc 2} & \multicolumn{2}{c}{True Tox. Prob.} & 0.11  & 0.19  & \textbf{0.31} & 0.39  & 0.46 \\
          & \multicolumn{2}{c}{Utility} & 0.428 & 0.432 & \textbf{0.438} & 0.442 & 0.438 \\
          & \multirow{3}[0]{*}{PEDOOP} & \# of Patients & 5.64  & 7.728 & \textbf{8.217} & 4.725 & 3.588 \\
          &       & Admissible Doses  & 2.8   & 14.5  & \textbf{24.4} & 12.5  & 5.2 \\
          &       & OBD  & 1     & 8.6   & \textbf{16.4} & 6.6   & 2.4 \\
          & \multirow{3}[1]{*}{DOOP} & \# of Patients & 5.523 & 7.728 & \textbf{8.178} & 4.512 & 4.008 \\
          &       & Admissible Doses  & 36.6  & 36.9  & \textbf{24.2} & 11.5  & 3.7 \\
          &       & OBD  & 13.7  & 15.8  & \textbf{12.3} & 6.4   & 1.4 \\
    \hline
    \multirow{8}[2]{*}{Sc 3} & \multicolumn{2}{c}{True Tox. Prob.} & 0.04  & 0.09  & 0.2   & \textbf{0.3} & 0.39 \\
          & \multicolumn{2}{c}{Utility} & 0.456 & 0.472 & 0.482 & \textbf{0.478} & 0.466 \\
          & \multirow{3}[0]{*}{PEDOOP} & \# of Patients & 3.492 & 3.99  & 6.78  & \textbf{7.017} & 8.721 \\
          &       & Admissible Doses  & 2.8   & 21.5  & 47.4  & \textbf{35.7} & 18.5 \\
          &       & OBD  & 0.7   & 9     & 28.1  & \textbf{17.7} & 7.8 \\
          & \multirow{3}[1]{*}{DOOP} & \# of Patients & 3.606 & 4.062 & 6.786 & \textbf{6.45} & 9.069 \\
          &       & Admissible Doses  & 39.6  & 45.6  & 51.8  & \textbf{33.2} & 13 \\
          &       & OBD  & 10.7  & 17.3  & 23.6  & \textbf{14.4} & 5.8 \\
    \hline
    \multirow{8}[2]{*}{Sc 4} & \multicolumn{2}{c}{True Tox. Prob.} & 0.11  & 0.16  & 0.21  & 0.24  & \textbf{0.27} \\
          & \multicolumn{2}{c}{Utility} & 0.428 & 0.444 & 0.478 & 0.502 & \textbf{0.514} \\
          & \multirow{3}[0]{*}{PEDOOP} & \# of Patients & 5.388 & 5.325 & 5.34  & 4.077 & \textbf{9.792} \\
          &       & Admissible Doses  & 2     & 15.3  & 43.1  & 47    & \textbf{39.3} \\
          &       & OBD  & 0.4   & 6     & 18.8  & 22.3  & \textbf{16} \\
          & \multirow{3}[1]{*}{DOOP} & \# of Patients & 5.244 & 4.971 & 5.367 & 4.227 & \textbf{10.116} \\
          &       & Admissible Doses  & 33.4  & 38    & 44.5  & 41.5  & \textbf{28.7} \\
          &       & OBD  & 8.6   & 15    & 17.2  & 18    & \textbf{11.6} \\
    \hline
    \end{tabular}%
  \label{tab:sim1}%
\end{table}%

\subsubsection{Sensitivity Analysis} 
 We conduct a small sensitivity analysis to check the robustness of model misspecification, especially on the PK modeling.  %To check if the generative model of $X_{ij}$ affects the performance of the proposed model remarkably, w
We modify the generative model in \eqref{eq:true_AUC1} and assume a new model given by  
\begin{equation} \label{eq:true_AUC_sen}
V_i \sim Gamma(4,1), \quad k_i \sim 1 + Gamma(3,1), \quad \log(X_{ij}) \mid V_i, k_i  \sim N\left(\frac{d_i}{V_i} - k_i^{e_i} t_j,1\right),
\end{equation}
where $e_i \sim Beta(1,d_i)$. Since $k_i > 1$, and $e_i$ decreases with $d_i$ in probability, \eqref{eq:true_AUC_sen} allows the patient-specific elimination rate $k_i^{e_i}$ to be lower for a higher dose.  The generative model \eqref{eq:true_AUC_sen} is now different from the true model \eqref{eq:true_AUC1} for PEDOOP. We re-run simulations with the new generative model of $X_{ij}$ %following the setup in Section \ref{sec:sim1_setup}. 
and summarize the new simulation results  in Table \ref{tab:sim1_sensitivity}. We call the new approach ``PEDOOP$-$w" with ``w" standing for the wrong fitting model. It can be seen that the performance of PEDOOP$-$w drops in scenario 1 selecting the  non-promising doses 1 and 2  with much higher frequencies. For other scenarios, PEDOOP$-$w performs similarly to PEDOOP.  These results suggest that when the PK-model is misspecified, the performance of PEDOOP is somewhat robust (as seen in Scenarios 2-4). However, in Scenario 1 when all doses are either too toxic or not effective, the model misspecification has more impact and reduced the performance of the design.    %did not change a lot except for scenario 1. In terms of ``Sel \% with U", PEDOOP remains unchanged, especially in scenarios 3 and 4, which shows that PEDOOP is stable.

\begin{table}[!htbp]
  \centering
  \renewcommand{\arraystretch}{0.7}
  \caption{ Sensitivity analysis comparing the proposed PEDOOP design with the same design but using a wrong model (PEDOOP$-$w) for generating the simulated data.  The bold font indicates the true OBD  in each scenario, which has  a  toxicity rate closest to $p_T = 0.3$ and an efficacy rate  larger than $\pi_E = 0.2$. PEDOOP represents the simulation results given the generative model in \eqref{eq:true_AUC1}, while ``PEDOOP$-$w" represents those given the ``wrong" generative model in \eqref{eq:true_AUC_sen}. For simplicity, we only present frequencies of selecting the OBD and omit the ones for Admissible dose.  }
    \begin{tabular}{c|c|llllll}
          & \multicolumn{2}{c}{Dose} & 1     & 2     & 3     & 4     & 5 \\
          & \multicolumn{2}{c}{True Eff. Prob.} & 0.12  & 0.18  & 0.27  & 0.33  & 0.37 \\
    \hline
    \multirow{6}[2]{*}{Sc 1} & \multicolumn{2}{c}{True Tox. Prob.} & 0.17  & 0.29  & 0.45  & 0.55  & 0.62 \\
          & \multicolumn{2}{c}{Utility} & 0.404 & 0.392 & 0.382 & 0.378 & 0.374 \\
          & \multirow{2}[0]{*}{PEDOOP} & \# of Patients & 10.494 & 11.163 & 5.79  & 1.587 & 0.627 \\
          &       & OBD  & 1     & 5.6   & 2.8   & 0     & 0.1 \\
          & \multirow{2}[1]{*}{PEDOOP-w} & \# of Patients & 10.719 & 10.167 & 6.093 & 1.731 & 0.9 \\
          &       & OBD  & 11.9  & 8     & 2.2   & 0.3   & 0.1 \\
    \hline
    \multirow{6}[2]{*}{Sc 2} & \multicolumn{2}{c}{True Tox. Prob.} & 0.11  & 0.19  & \textbf{0.31} & 0.39  & 0.46 \\
          & \multicolumn{2}{c}{Utility} & 0.428 & 0.432 & \textbf{0.438} & 0.442 & 0.438 \\
          & \multirow{2}[0]{*}{PEDOOP} & \# of Patients & 5.64  & 7.728 & \textbf{8.217} & 4.725 & 3.588 \\
          &       & OBD  & 1     & 8.6   & \textbf{16.4} & 6.6   & 2.4 \\
          & \multirow{2}[1]{*}{PEDOOP-w} & \# of Patients & 5.745 & 6.894 & \textbf{7.914} & 4.296 & 5.049 \\
          &       & OBD  & 5.8   & 12.8  & \textbf{16.2} & 6.9   & 2.5 \\
    \hline
    \multirow{6}[2]{*}{Sc 3} & \multicolumn{2}{c}{True Tox. Prob.} & 0.04  & 0.09  & \textbf{0.2} & 0.3   & 0.39 \\
          & \multicolumn{2}{c}{Utility} & 0.456 & 0.472 & \textbf{0.482} & 0.478 & 0.466 \\
          & \multirow{2}[0]{*}{PEDOOP} & \# of Patients & 3.492 & 3.99  & \textbf{6.78} & 7.017 & 8.721 \\
          &       & OBD  & 0.7   & 9     & \textbf{28.1} & 17.7  & 7.8 \\
          & \multirow{2}[1]{*}{PEDOOP-w} & \# of Patients & 3.579 & 4.011 & \textbf{6.594} & 5.628 & 10.188 \\
          &       & OBD  & 2     & 8.4   & \textbf{29.2} & 18    & 8.8 \\
    \hline
    \multirow{6}[2]{*}{Sc 4} & \multicolumn{2}{c}{True Tox. Prob.} & 0.11  & 0.16  & 0.21  & 0.24  & \textbf{0.27} \\
          & \multicolumn{2}{c}{Utility} & 0.428 & 0.444 & 0.478 & 0.502 & \textbf{0.514} \\
          & \multirow{2}[0]{*}{PEDOOP} & \# of Patients & 5.388 & 5.325 & 5.34  & 4.077 & \textbf{9.792} \\
          &       & OBD  & 0.4   & 6     & 18.8  & 22.3  & \textbf{16} \\
          & \multirow{2}[1]{*}{PEDOOP-w} & \# of Patients & 5.133 & 4.608 & 4.959 & 3.747 & \textbf{11.529} \\
          &       & OBD  & 2.5   & 5.6   & 18.1  & 26.5  & \textbf{18.9} \\
    \hline
    \end{tabular}%
  \label{tab:sim1_sensitivity}%
\end{table}%

\subsection{Simulation 2}
\subsubsection{Simulation Setup} 
We perform 1,000 simulation studies to compare the proposed PEDOOP design and the SEARS design \citep{pan2014phase}. In phase I, similarly we assume that there are seven dose levels, (15, 30, 60, 90, 120), and drug concentration are measured for each patient six times, at 1, 3, 5, 7, 12, and 24 hours after administration.
A maximum of 30 patients with three patients  a cohort is to be enrolled in phase I. The maximum sample size of phase II  is set to 150 with 10 patients  a cohort.   This means that the BAR randomization probability is updated every 10 patients in phase II.    Similarly, the maximum sample size of phase I for the SEARS design is set to 30, and the maximum sample size of the both phases under the SEARS design is set to 180. Besides, the SEARS design considered a maximum sample size  of 36 patients  for each dose and the control arm. %, which didn't allow more than 36 patients being treated at a dose or the control arm. 
For both designs, $\pi_T = 0.2$, $p^* = 0.6$, $q^* = 0.6$, and $\pi_E = 0.2$ in the graduation rule \eqref{eq:gra_rule} at the end of phase I, and $p^{**} = 0.7$ and $q^{**} = 0.9$ in the selection criteria \eqref{eq:sel_rule} at the end of phase II.

In phase I, we generate the patient-specific elimination rate, $k$, volume of distribution $V$, and drug concentration data $X$ for 30 individuals randomly via 
\begin{equation} \label{eq:true_AUC2}
V_i \sim Gamma(10,1), \quad k_i \sim Gamma(9,1.5), \quad \log(X_{ij}) \mid V_i, k_i  \sim N\left(\log\left(\frac{d_i}{V_i}\right) - k_i t_j,4^2\right),
\end{equation}
where $d_i \in \{15, 30, 60, 90, 120\}$ and $t_j \in \{1, 3, 5, 7, 12, 24\}$.

We set up a total of 12 scenarios in the simulation study. All the scenarios have the same control arm with a toxicity probability of 0.17 and an efficacy probability of 0.2. Following \cite{pan2014phase}, we only consider two types of toxicity profile. That is, scenarios 1 to 6 are with toxicity probabilities $(0.03,0.06,0.17,0.3,0.5)$, which means that the third dose of the tested drug has equal toxicity as the control arm, and the fourth one is more toxic than the control arm. Scenarios 7 to 12 are with toxicity probabilities $(0.03,0.06,0.09,0.12,0.15)$, where any dose is with less toxic than the control arm. 

In scenarios 1 and 7, the efficacy probability stays flat regardless of the dose. Therefore, the first dose is the optimal one in the two scenarios. 
The efficacy probability in scenarios 2 and 8 increases with dose , while in contrast efficacy probability decreases with dose in scenarios 3 and 9.
Scenarios 4 and 10 have a N-shape efficacy probability and scenarios 5 and 11 have a U-shape efficacy probability, although the two types are not common. Efficacy probability in scenarios 6 and 12 reaches a plateau and stays flat after a certain dose level. Specific information can be found in Tables \ref{tab:sim_sc1_6} and \ref{tab:sim_sc7_12}.

% We drew 10000 posterior samples, with the first 5000 being burn-in and the rest being used for posterior inference. We assessed convergence by visually examining the trace plots. 
% Figure \ref{fig:posterior} is a trace plot of an example of the data analysis for one completed simulated trail. The trace plots suggest that almost all of the parameters are convergence. 

\subsubsection{Simulation Results} 
The operating characteristics of the PEDOOP and SEARS designs are summarized in Tables \ref{tab:sim_sc1_6} and \ref{tab:sim_sc7_12}. %Below we first discuss the comparison of PEDOOP and SEARS on ``\# of Patients" and ``Sel \%", which examine the patient allocation and the dose selection of the designs. 
In this simulation study, ``Admissible Doses" is presented to indicate the selection frequency of each dose at the end of phase II according to the criteria \eqref{eq:sel_rule}. This rule may lead to the selection of multiple doses. Also, we present the ``OBD" is based on the criteria \eqref{eq:sel_rule} and the highest posterior probability \eqref{eq:prob_maxu}, where a maximum of one dose is selected.

In scenarios 1 and 7, there are no desirable doses since all doses of the tested drug have the same efficacy probability as the control arm. The PEDOOP design stop trials with fewer enrolled patients, % earlier than the SEARS design, enrolled fewer patients, and selected doses in a smaller proportion. For example, PEDOOP allocated 
allocating on average  9.877 and 12.228 patients to doses 2 and 3 in scenario 1, respectively, compared with  18.483 and 19.428 patients by SEARS. PEDOOP also selects doses with less frequencies, which is as expected since no doses should be selected under the two scenarios. %selects doses 2 and 3 with a total proportion of 0.045, while SEARS selected doses 2 and 3 with a total proportion of 0.097.

In scenarios 2 and 6,  the optimal dose for both scenarios is dose 3. %of the tested drug. 
Dose 2 is the second optimal with a similar utility. %the utility of dose 2 is close to that of dose 3. 
PEDOOP and SEARS select doses 2 and 3 with similar frequencies, but SEARS tends to allocate more patients to toxic doses, e.g., doses 4 and 5, than PEDOOP. %For example, PEDOOP allocated 7.747 and 7.534 patients at dose 4 on average in scenarios 2 and 6, respectively, while SEARS allocated 16.164 and 13.68 patients.

In scenarios 8 and 12, none of the doses are toxic as their toxicity probabilities are all below $p_T$. Therefore, the optimal dose is the one with the highest %depends on the 
efficacy probability. In scenario 8, the efficacy probability increases with the dose level, and dose 5 is optimal. In scenario 12, the efficacy probability reaches a plateau after dose 3, making it the optimal dose. PEDOOP and SEARS allocate similar numbers of patients to the doses while  PEDOOP selects the optimal dose with a lower proportion than SEARS.

In scenarios 3, 5, 9, and 11, the optimal dose is dose 1 for all of them based on the truth. %, due to the specific assumptions about the efficacy probability. 
In terms of patient allocation, PEDOOP perform similar to SEARS. However, PEDOOP selects the optimal dose with less frequencies. This is potentially due to the fact that efficacy does not increase with dose in these scenarios, which violates the monotonic efficacy assumption in \eqref{eq:q_d}. %, which implies the efficacy probability increases with the PD drug effect, as well as with the drug concentration and the dose level.

In scenarios 4 and 10, efficacy probabilities of doses display an ``n" shape. Dose 3 has the highest efficacy, and it is the optimal dose. In terms of patient allocation, PEDOOP is safer than SEARS in scenario 4 as it allocates more patients to doses 1 and 2 and fewer patients to doses 3 to 5 than SEARS.  In scenario 10,  all doses of the tested drug are safe, and PEDOOP  tends to allocate more patients to dose 5. Both designs select the optimal dose in a similar fashion in both scenarios. % select dose 2 with a higher frequency than dose 3 in scenario 4. This may be because the toxicity probability of dose 3 is close to $p_T$, and it sometimes results in data that doesn't meet the toxicity condition of \eqref{eq:sel_rule}. Also, if all doses of the tested drug are safe, like in scenario 10, both designs selected dose 3 with the highest probability. So, in terms of selection proportion, PEDOOP is similar to SEARS.

In most scenarios, ``OBD" of the optimal dose did not decrease significantly compared to ``Admissible Doses". However, in scenarios 3, 11, and 12, ``OBD" of the optimal dose dropped considerably. This is primarily because other doses also have high utilities, similar to the optimal one. For example, in scenario 12, dose 3 has the highest utility of 0.664, while doses 4 and 5 are with 0.652 and 0.64. Therefore, it is difficult to differentiate between these doses.

\begin{table}[!htbp]
  \centering
  \renewcommand{\arraystretch}{0.6}
  \caption{Simulation results of scenarios 1 to 6 in Simulation 2.}
    \begin{tabular}{c|c|lllllll}
          & \multicolumn{2}{c}{Dose} & 1     & 2     & 3     & 4     & 5     & Control \\
          & \multicolumn{2}{c}{True Tox. Prob.} & 0.03  & 0.06  & 0.17  & 0.3   & 0.5   & 0.17 \\
          & \multicolumn{2}{c}{AUC} & 0.31  & 0.62  & 1.25  & 1.88  & 2.5   &  \\
    \hline
    \multirow{7}[2]{*}{Sc 1} & \multicolumn{2}{c}{True Eff. Prob.} & 0.2   & 0.2   & 0.2   & 0.2   & 0.2   & 0.2 \\
          & \multicolumn{2}{c}{Utility} & 0.508 & 0.496 & 0.452 & 0.4   & 0.32  & 0.452 \\
          & \multirow{3}[0]{*}{PEDOOP} & \# of Patients & 5.519 & 9.877 & 12.228 & 6.049 & 2.818 & 3.558 \\
          &       & Admissible Doses & 0.3   & 1.3   & 1     & 0     & 0     &  \\
          &       & OBD & 0.3   & 1.2   & 1     & 0     & 0     &  \\
          & \multirow{2}[1]{*}{SEARS} & \# of Patients & 16.968 & 18.483 & 19.428 & 11.652 & 2.982 & 14.304 \\
          &       & Admissible Doses & 7.3   & 6.5   & 3.2   & 0.2   & 0     &  \\
    \hline
    \multirow{7}[2]{*}{Sc 2} & \multicolumn{2}{c}{True Eff. Prob.} & 0.2   & 0.3   & 0.5   & 0.7   & 0.8   & 0.2 \\
          & \multicolumn{2}{c}{Utility} & 0.508 & 0.556 & 0.632 & 0.7   & 0.68  & 0.452 \\
          & \multirow{3}[0]{*}{PEDOOP} & \# of Patients & 17.826 & 25.598 & 22.326 & 7.518 & 3.116 & 14.622 \\
          &       & Admissible Doses & 7.4   & 37.9  & 27.9  & 0.8   & 0     &  \\
          &       & OBD & 4     & 29.5  & 25    & 0.8   & 0     &  \\
          & \multirow{2}[1]{*}{SEARS} & \# of Patients & 12.216 & 17.349 & 25.176 & 16.164 & 3.561 & 10.293 \\
          &       & Admissible Doses & 10.4  & 35.4  & 34.7  & 1.8   & 0     &  \\
    \hline
    \multirow{7}[2]{*}{Sc 3} & \multicolumn{2}{c}{True Eff. Prob.} & 0.8   & 0.7   & 0.5   & 0.3   & 0.2   & 0.2 \\
          & \multicolumn{2}{c}{Utility} & 0.868 & 0.796 & 0.632 & 0.46  & 0.32  & 0.452 \\
          & \multirow{3}[0]{*}{PEDOOP} & \# of Patients & 32.189 & 25.086 & 14.988 & 6.322 & 2.838 & 5.325 \\
          &       & Admissible Doses & 90.9  & 81.3  & 25.2  & 0.2   & 0     &  \\
          &       & OBD & 76.1  & 19.5  & 0.6   & 0     & 0     &  \\
          & \multirow{2}[1]{*}{SEARS} & \# of Patients & 30.039 & 23.022 & 15.153 & 8.535 & 2.922 & 5.754 \\
          &       & Admissible Doses & 94.9  & 85.1  & 28.7  & 1.9   & 0     &  \\
    \hline
    \multirow{7}[2]{*}{Sc 4} & \multicolumn{2}{c}{True Eff. Prob.} & 0.2   & 0.4   & 0.8   & 0.4   & 0.2   & 0.2 \\
          & \multicolumn{2}{c}{Utility} & 0.508 & 0.616 & 0.812 & 0.52  & 0.32  & 0.452 \\
          & \multirow{3}[0]{*}{PEDOOP} & \# of Patients & 14.372 & 23.361 & 24.709 & 6.432 & 2.839 & 11.146 \\
          &       & Admissible Doses & 6.6   & 57.7  & 28.5  & 0.9   & 0     &  \\
          &       & OBD & 2.3   & 41.6  & 28.5  & 0.1   & 0     &  \\
          & \multirow{2}[1]{*}{SEARS} & \# of Patients & 9.435 & 15.165 & 30.399 & 8.913 & 2.646 & 7.032 \\
          &       & Admissible Doses & 8.2   & 52.6  & 37.9  & 3.4   & 0.1   &  \\
    \hline
    \multirow{7}[2]{*}{Sc 5} & \multicolumn{2}{c}{True Eff. Prob.} & 0.8   & 0.4   & 0.2   & 0.4   & 0.8   & 0.2 \\
          & \multicolumn{2}{c}{Utility} & 0.868 & 0.616 & 0.452 & 0.52  & 0.68  & 0.452 \\
          & \multirow{3}[0]{*}{PEDOOP} & \# of Patients & 31.85 & 15.688 & 14.06 & 6.436 & 2.848 & 6.35 \\
          &       & Admissible Doses & 75.8  & 46.8  & 2.9   & 0.9   & 0     &  \\
          &       & OBD & 75.8  & 8.2   & 0     & 0     & 0     &  \\
          & \multirow{2}[1]{*}{SEARS} & \# of Patients & 31.851 & 12.351 & 11.739 & 9.522 & 3.456 & 6.318 \\
          &       & Admissible Doses & 88.5  & 44.4  & 4.5   & 2.1   & 0     &  \\
    \hline
    \multirow{7}[2]{*}{Sc 6} & \multicolumn{2}{c}{True Eff. Prob.} & 0.2   & 0.4   & 0.5   & 0.5   & 0.5   & 0.2 \\
          & \multicolumn{2}{c}{Utility} & 0.508 & 0.616 & 0.632 & 0.58  & 0.5   & 0.452 \\
          & \multirow{3}[0]{*}{PEDOOP} & \# of Patients & 16.195 & 29.79 & 21.727 & 6.878 & 3.041 & 12.047 \\
          &       & Admissible Doses & 7.4   & 69.5  & 29.9  & 0.9   & 0     &  \\
          &       & OBD & 2.7   & 55.7  & 21    & 0.6   & 0     &  \\
          & \multirow{2}[1]{*}{SEARS} & \# of Patients & 12.162 & 21.708 & 26.586 & 13.68 & 3.519 & 10.491 \\
          &       & Admissible Doses & 8.6   & 66.8  & 37    & 1.8   & 0     &  \\
    \hline
    \end{tabular}%
  \label{tab:sim_sc1_6}%
\end{table}

\begin{table}[htbp]
  \centering
  \renewcommand{\arraystretch}{0.6}
  \caption{Simulation results of scenarios 7 to 12 in Simulation 2. }
    \begin{tabular}{c|c|lllllll}
          & \multicolumn{2}{c}{Dose} & 1     & 2     & 3     & 4     & 5     & Control \\
          & \multicolumn{2}{c}{True Tox. Prob.} & 0.03  & 0.06  & 0.09  & 0.12  & 0.15  & 0.17 \\
          & \multicolumn{2}{c}{AUC} & 0.31  & 0.62  & 1.25  & 1.88  & 2.5   &  \\
    \hline
    \multirow{7}[2]{*}{Sc 7} & \multicolumn{2}{c}{True Eff. Prob.} & 0.2   & 0.2   & 0.2   & 0.2   & 0.2   & 0.2 \\
          & \multicolumn{2}{c}{Utility} & 0.508 & 0.496 & 0.484 & 0.472 & 0.46  & 0.452 \\
          & \multirow{3}[0]{*}{PEDOOP} & \# of Patients & 4.713 & 6.416 & 7.803 & 7.45  & 15.088 & 4.426 \\
          &       & Admissible Doses & 0.4   & 0.9   & 1.9   & 2.2   & 1.8   &  \\
          &       & OBD & 0.4   & 0.9   & 1.6   & 1.9   & 1.6   &  \\
          & \multirow{2}[1]{*}{SEARS} & \# of Patients & 16.86 & 16.668 & 17.475 & 18.054 & 17.916 & 15.189 \\
          &       & Admissible Doses & 5.9   & 6.8   & 6.3   & 5.4   & 4.9   &  \\
    \hline
    \multirow{7}[2]{*}{Sc 8} & \multicolumn{2}{c}{True Eff. Prob.} & 0.2   & 0.3   & 0.5   & 0.7   & 0.8   & 0.2 \\
          & \multicolumn{2}{c}{Utility} & 0.508 & 0.556 & 0.664 & 0.772 & 0.82  & 0.452 \\
          & \multirow{3}[0]{*}{PEDOOP} & \# of Patients & 7.456 & 10.085 & 14.034 & 18.842 & 26.272 & 8.6 \\
          &       & Admissible Doses & 4.5   & 18.4  & 40.1  & 45.1  & 42.7  &  \\
          &       & OBD & 1.1   & 6.3   & 13.4  & 27.1  & 37.8  &  \\
          & \multirow{2}[1]{*}{SEARS} & \# of Patients & 7.683 & 10.164 & 14.748 & 22.194 & 25.788 & 6.537 \\
          &       & Admissible Doses & 7.9   & 21    & 54.9  & 58.2  & 43.9  &  \\
    \hline
    \multirow{7}[2]{*}{Sc 9} & \multicolumn{2}{c}{True Eff. Prob.} & 0.8   & 0.7   & 0.5   & 0.3   & 0.2   & 0.2 \\
          & \multicolumn{2}{c}{Utility} & 0.868 & 0.796 & 0.664 & 0.532 & 0.46  & 0.452 \\
          & \multirow{3}[0]{*}{PEDOOP} & \# of Patients & 26.843 & 20.797 & 13.631 & 8.349 & 13.704 & 6.392 \\
          &       & Admissible Doses & 70.8  & 65.6  & 42.3  & 9.6   & 0.7   &  \\
          &       & OBD & 61    & 19.6  & 7.1   & 1.3   & 0.1   &  \\
          & \multirow{2}[1]{*}{SEARS} & \# of Patients & 27.81 & 20.925 & 12.942 & 9.867 & 9.483 & 6.048 \\
          &       & Admissible Doses & 86.5  & 75.5  & 45.4  & 14.6  & 4.2   &  \\
    \hline
    \multirow{7}[2]{*}{Sc 10} & \multicolumn{2}{c}{True Eff. Prob.} & 0.2   & 0.4   & 0.8   & 0.4   & 0.2   & 0.2 \\
          & \multicolumn{2}{c}{Utility} & 0.508 & 0.616 & 0.844 & 0.592 & 0.46  & 0.452 \\
          & \multirow{3}[0]{*}{PEDOOP} & \# of Patients & 6.267 & 11.257 & 27.68 & 10.503 & 14.708 & 6.811 \\
          &       & Admissible Doses & 3     & 31.9  & 60.9  & 21.5  & 1     &  \\
          &       & OBD & 0.1   & 7.5   & 60.6  & 5.4   & 0.4   &  \\
          & \multirow{2}[1]{*}{SEARS} & \# of Patients & 7.881 & 11.847 & 30.69 & 11.46 & 9.27  & 6.465 \\
          &       & Admissible Doses & 7.8   & 42.5  & 77.2  & 27.2  & 3.9   &  \\
    \hline
    \multirow{7}[2]{*}{Sc 11} & \multicolumn{2}{c}{True Eff. Prob.} & 0.8   & 0.4   & 0.2   & 0.4   & 0.8   & 0.2 \\
          & \multicolumn{2}{c}{Utility} & 0.868 & 0.616 & 0.484 & 0.592 & 0.82  & 0.452 \\
          & \multirow{3}[0]{*}{PEDOOP} & \# of Patients & 31.639 & 9.546 & 6.451 & 7.257 & 24.252 & 5.948 \\
          &       & Admissible Doses & 90    & 30.7  & 2.5   & 12.7  & 41.3  &  \\
          &       & OBD & 75.4  & 2     & 0     & 0.8   & 17.7  &  \\
          & \multirow{2}[1]{*}{SEARS} & \# of Patients & 25.194 & 9.927 & 9.18  & 10.95 & 23.583 & 5.706 \\
          &       & Admissible Doses & 86.4  & 37.9  & 6.9   & 27.7  & 43.5  &  \\
    \hline
    \multirow{7}[2]{*}{Sc 12} & \multicolumn{2}{c}{True Eff. Prob.} & 0.2   & 0.4   & 0.5   & 0.5   & 0.5   & 0.2 \\
          & \multicolumn{2}{c}{Utility} & 0.508 & 0.616 & 0.664 & 0.652 & 0.64  & 0.452 \\
          & \multirow{3}[0]{*}{PEDOOP} & \# of Patients & 8.916 & 17.566 & 22.636 & 18.463 & 19.22 & 10.558 \\
          &       & Admissible Doses  & 4.5   & 46.7  & 55.1  & 41.6  & 23.8  &  \\
          &       & OBD  & 0.6   & 18.3  & 34.6  & 24.7  & 13    &  \\
          & \multirow{2}[1]{*}{SEARS} & \# of Patients & 10.635 & 18.051 & 23.328 & 22.206 & 21.315 & 9.432 \\
          &       & Admissible Doses  & 11.3  & 58    & 70.8  & 56.2  & 39.6  &  \\
    \hline
    \end{tabular}%
  \label{tab:sim_sc7_12}%
\end{table}%

\clearpage
\newpage
\section{Discussion}\label{sec:discussion}
Phase I trials typically collect data on toxicity, efficacy, PK, and PD of a new drug. However, most existing phase I dose-finding designs only focus on toxicity and efficacy outcomes, ignoring the PK and PD information. This may lead to inconsistency between the MTD identified by these designs and the optimal dose suggested by the PK and PD data. In this paper, we propose a statistical model that incorporates toxicity, efficacy, PK, and PD outcomes meanwhile to facilitate better dose selection in phase I trials, which is also the main contribution of this paper.
A novel feature of the model is the use of dose-level AUC and PD, rather than patient-level ones, as covariates for toxicity and efficacy probabilities. Another contribution of this paper is the simulation of phase II trials to demonstrate the practicality and advantages of the proposed model. 

In the proposed model, we assume that toxicity and efficacy increase monotonically with the dose level. However, in real trials, there may be various possible scenarios. The model may need to be adjusted according to the actual drug properties.

 In phase II, we set a cohort size of 10 to match that in   the SEARS design \citep{pan2014phase}.  In practical trials, the choice of the optimal cohort size  should take into account  the enrollment speed, followup time for efficacy and toxicity outcomes, trial duration, logistic cost and burden, and benefits from adjusting randomization probabilities. It is worth more consideration for future work.  %maximizing data availability and timely patient benefit, taking into account the actual observation time for efficacy and toxicity outcomes and the expected patient benefit in trials.

The proposed PEDOOP design  enrolls patients in cohorts in phase II. After patients in a cohort are randomized at doses graduated from phase I or the control arm, they are followed for a period of time for toxicity and efficacy outcomes. Then, the randomization probabilities are updated based on the posterior distribution of the proposed utilities of arms in Section \ref{sec:utility}, which are only based on toxicity and efficacy outcomes. 
We did not consider PK/PD in Phase II simply because they might not be available in time in practical situations as these measurements usually take time to produce in real life.  We assume that the Phase II design would only use toxicity and efficacy outcomes for calculating patients
allocation probabilities in the proposed BAR procedure, but at the end, a dose selection and go/no-go decision will be made based on all the data, including clinical and PK/PD data.

Finally, the proposed model is based on a first-order one-compartment model for the IV injection. This is a basic model in PK models. In future research, the proposed model can be extended to more situations. For example, the PK model we used leads to a linear relationship between AUC and dose which may not hold for certain therapeutics like monoclonal antibodies (mAbs) \citep{tabrizi2006elimination}. This is because binding of the drug to the target becomes a clearance mechanism, which may lead to potential alterations in clearance, particularly at lower doses. As the dose increases, the target-mediated clearance becomes saturated, and drug exposure may then become linear. 
Moreover, we introduce the concept of drug effect density \eqref{eq:r_d} in the proposed model to link drug concentration with PD and efficacy. This can also be realized by replacing the population-level drug concentration $c(t \mid d)$ with the area under the concentration-time curve $AUC(d)$. To improve the practical utility of our models, we may further elaborate on this adjustment and discuss how it can enhance precision in capturing drug-response relationships, as well as a more sophisticated Emax model, outlined in $r(d) = E_0 + \frac{\emax \times (AUC(d))^{\gamma}}{\edft^{\gamma} + (AUC(d))^{\gamma}}$.

\section*{Acknowledgement}
We thank Michael J. Fossler for providing insightful comments. 

\newpage

\bibliographystyle{apalike}
\bibliography{PEDOOP.bib}

\newpage

\begin{appendices}
\appendixpage
\setcounter{table}{0}
\renewcommand{\thetable}{A.\arabic{table}}
\setcounter{equation}{0}
\renewcommand{\theequation}{A.\arabic{equation}}
\setcounter{figure}{0}
\renewcommand{\thefigure}{A.\arabic{figure}}
\section{Prior on $\phi$} \label{appendix:prior_phi}
\subsection{Hypergeometric Function}
For any value $\phi > 1$, there is a closed form for \eqref{eq:eta_d} theoretically \citep[Ch. 15]{abramowitz1968handbook}, which can be represented by
\begin{equation*}
    \eta(d) = C(d)\emax \int_{\lambda_k}^{\infty} \frac{1}{x^{\phi} + C(d)} dx = C(d)\emax  \times \left(\frac{x}{C(d)} F(1,\frac{1}{\phi};1+\frac{1}{\phi};-\frac{x^{\phi}}{C(d)})\right)\bigg|_{\lambda_k}^{\infty}.
\end{equation*}
Here, $F(a,b;c;z)$ denotes the Gaussian hypergeometric function.
\begin{equation} \label{eq:F}
    F(a,b;c;z) = \sum_{n=0}^\infty \frac{(a)_n (b)_n}{(c)_n} \frac{z^n}{n!} = 1 + \frac{ab}{c}\frac{z}{1!} + \frac{a(a+1)b(b+1)}{c(c+1)}\frac{z^2}{2!} + \cdots.
\end{equation}
An integral representation of \eqref{eq:F} is 
\begin{equation*}
    F(a,b;c;z) = \frac{\Gamma(c)}{\Gamma(b)\Gamma(c-b)} \int_{0}^{1} t^{b-1}(1-t)^{c-b-1}(1-tz)^{-a} dt, \quad c > b > 0.
\end{equation*}
Therefore, we can find
\begin{equation*}
    \begin{aligned}
        \frac{\partial}{\partial x}\left(\frac{x}{C(d)} F(1,\frac{1}{\phi};1+\frac{1}{\phi};-\frac{x^{\phi}}{C(d)})\right) & = \frac{\partial}{\partial x} \left( \frac{x}{C(d)} \times \frac{1}{\phi} \int_0^1 t^{\frac{1}{\phi} - 1} (1+\frac{tx^{\phi}}{C(d)})^{-1} dt\right) \\
        & = \frac{\partial}{\partial x} \left( \frac{x}{C(d)} \times \int_0^1 \frac{C(d)}{C(d) + (t^{\frac{1}{\phi}}x)^{\phi}} dt^{\frac{1}{\phi}}\right) \\
        &\xlongequal{s=t^{\frac{1}{\phi}}x} \frac{\partial}{\partial x} \left( \int_0^x \frac{1}{C(d) + s^{\phi}} ds\right)\\
        & =   \frac{1}{x^{\phi} + C(d)} 
    \end{aligned}
\end{equation*}
By calling the functions in the GNU Scientific Library of computing the values of \eqref{eq:F}, the MCMC algorithm is achievable with a continuous prior distribution on $\phi$ or $\gamma$ in \eqref{eq:eta_d}. However, the computation of the hypergeometric function is difficult to implement in JAGS, by which our simulation code is written mainly.

\subsection{Categorical Prior}
For simplicity, we assume a categorical prior distribution on $\phi$, $\phi \sim \text{Cat}(4,\boldsymbol{p}),$
where $\phi \in \{2,3,4,5\}$ and $\boldsymbol{p} = (0.25,0.25,0.25,0.25)$. These values of $\phi$ result in distinct closed forms for $\eta(d)$.

When $\phi = 2$,
\begin{equation*}
\eta(d) = \sqrt{C(d)} \emax  \left[ \frac{\pi}{2} - \arctan\left(\frac{\lambda_k}{\sqrt{C(d)}}\right) \right].
\end{equation*}
When $\phi = 3$,
\begin{equation*}
	\begin{aligned}
		\eta(d)&= \frac{1}{6} C(d)^{\frac{1}{3}} \emax \times 
            \left[2\log\left(C(d)^{\frac{1}{3}} + x\right) - \log\left(C(d)^{\frac{2}{3}} - C(d)^{\frac{1}{3}}x + x^2\right) + 2\sqrt{3}\arctan\left(\frac{2x - C(d)^{\frac{1}{3}}}{\sqrt{3}C(d)^{\frac{1}{3}}}\right)\right] \Bigg|_{\lambda_k}^{\infty}\\
		&= \frac{1}{6} C(d)^{\frac{1}{3}} \emax \times 
            \left[\sqrt{3} \pi - 
            2\log\left(C(d)^{\frac{1}{3}} + \lambda_k\right) + \log\left(C(d)^{\frac{2}{3}} - C(d)^{\frac{1}{3}}\lambda_k + \lambda_k^2\right) - 2\sqrt{3}\arctan\left(\frac{2\lambda_k - C(d)^{\frac{1}{3}}}{\sqrt{3}C(d)^{\frac{1}{3}}}\right)\right]. 
	\end{aligned}
\end{equation*}
When $\phi = 4$,
\begin{equation*}
    \begin{aligned}
        \eta(d) = & \frac{1}{4 \sqrt{2}} C(d)^{\frac{1}{4}} \emax   \times \Bigg[\log\left( x^2 + \sqrt{2}C(d)^{\frac{1}{4}}x + \sqrt{C(d)} \right) - \log\left( x^2 - \sqrt{2}C(d)^{\frac{1}{4}}x + \sqrt{C(d)} \right)   \\ 
        &   \qquad\qquad\qquad\qquad + 2\arctan\left(\frac{\sqrt{2}x}{C(d)^{\frac{1}{4}}} - 1\right)+ 2\arctan\left(\frac{\sqrt{2}x}{C(d)^{\frac{1}{4}}} + 1\right)\Bigg]\Bigg|_{\lambda_k}^{\infty} \\
        = & \frac{1}{4 \sqrt{2}} C(d)^{\frac{1}{4}} \emax   \times \Bigg[ \pi - \log\left( \lambda_k^2 + \sqrt{2}C(d)^{\frac{1}{4}}\lambda_k + \sqrt{C(d)} \right) + \log\left( \lambda_k^2 - \sqrt{2}C(d)^{\frac{1}{4}}\lambda_k + \sqrt{C(d)} \right)  \\ 
        &  \qquad\qquad\qquad\qquad  - 2\arctan\left(\frac{\sqrt{2}\lambda_k}{C(d)^{\frac{1}{4}}} - 1\right)- 2\arctan\left(\frac{\sqrt{2}\lambda_k}{C(d)^{\frac{1}{4}}} + 1\right)\Bigg].
    \end{aligned}
\end{equation*}
When $\phi = 5$,
\begin{equation*}
    \begin{aligned}
        \eta(d) = & \frac{1}{20} C(d)^{\frac{1}{5}} \emax   \times \Bigg[(\sqrt{5} - 1)\log\left( 2x^2 + (\sqrt{5} - 1)C(d)^{\frac{1}{5}}x + 2C(d)^{\frac{2}{5}} \right)   \\
        &   \qquad\qquad\qquad\qquad - (\sqrt{5} + 1)\log\left( 2x^2 + (\sqrt{5} + 1)C(d)^{\frac{1}{5}}x + 2C(d)^{\frac{2}{5}} \right) + 4\log\left(C(d)^{\frac{1}{5}} + x\right) \\
        &   \qquad\qquad\qquad\qquad + 2\sqrt{10 + 2\sqrt{5}}\arctan\left(\frac{4x + (\sqrt{5}-1)C(d)^{\frac{1}{5}}}{\sqrt{10 + 2\sqrt{5}}C(d)^{\frac{1}{5}}}\right) \\ 
        &   \qquad\qquad\qquad\qquad + 2\sqrt{10 - 2\sqrt{5}}\arctan\left(\frac{4x - (\sqrt{5}+1)C(d)^{\frac{1}{5}}}{\sqrt{10 - 2\sqrt{5}}C(d)^{\frac{1}{5}}}\right)\Bigg]\Bigg|_{\lambda_k}^{\infty} \\
        = & \frac{1}{20} C(d)^{\frac{1}{5}} \emax   \times \Bigg[ \left(\sqrt{10 + 2\sqrt{5}} + \sqrt{10 - 2\sqrt{5}}\right) \pi - (\sqrt{5} - 1)\log\left( 2\lambda_k^2 + (\sqrt{5} - 1)C(d)^{\frac{1}{5}}\lambda_k + 2C(d)^{\frac{2}{5}} \right)   \\
        &   \qquad\qquad\qquad\qquad + (\sqrt{5} + 1)\log\left( 2\lambda_k^2 + (\sqrt{5} + 1)C(d)^{\frac{1}{5}}\lambda_k + 2C(d)^{\frac{2}{5}} \right) - 4\log\left(C(d)^{\frac{1}{5}} + \lambda_k\right) \\
        &   \qquad\qquad\qquad\qquad - 2\sqrt{10 + 2\sqrt{5}}\arctan\left(\frac{4\lambda_k + (\sqrt{5}-1)C(d)^{\frac{1}{5}}}{\sqrt{10 + 2\sqrt{5}}C(d)^{\frac{1}{5}}}\right) \\ 
        &   \qquad\qquad\qquad\qquad - 2\sqrt{10 - 2\sqrt{5}}\arctan\left(\frac{4\lambda_k - (\sqrt{5}+1)C(d)^{\frac{1}{5}}}{\sqrt{10 - 2\sqrt{5}}C(d)^{\frac{1}{5}}}\right)\Bigg].
    \end{aligned}
\end{equation*}

After collecting data $\mathcal{D}_{\text{I}}$ from Phase I, the posterior distributions of $\phi$ in the four scenarios of Simulation 1 (Section \ref{sec:sim1}) are reported in Table \ref{tab:pos_phi}.

\begin{table}[!htbp]
  \centering
  \caption{Average posterior probabilities of $\phi$ in the four scenarios of Simulation 1,  averaged across 1,000 simulated trials.  ``Average MAP" denotes the average posterior mode of $\phi$ across 1,000 simulations. The true $\phi$ value is 2 in the simulation. }%And $\phi = x$ represents the average posterior probability $\Pr(\phi = x \mid \mathcal{D}_{\text{I}})$ across 1000 simulations, $x = 2,3,4,5$.}
    \begin{tabular}{c|llll|c}
          & \multicolumn{4}{c|}{Average $\Pr(\phi = x \mid \mathcal{D}_{\text{I}})$}  & Average \\
\cline{2-5}           & 2     & 3     & 4     & 5 & MAP \\
    \hline
    Sc 1   & 0.403 & 0.232 & 0.129 & 0.235 & 2.887\\
    Sc 2   & 0.47  & 0.237 & 0.116 & 0.177 & 2.467\\
    Sc 3   & 0.525 & 0.239 & 0.105 & 0.131 & 2.277\\
    Sc 4   & 0.504 & 0.235 & 0.11  & 0.151 & 2.367\\
    \hline
    \end{tabular}%
  \label{tab:pos_phi}%
\end{table}%

\end{appendices}

\end{document}